\documentclass[journal]{IEEEtran}

\usepackage[utf8]{inputenc} 
\usepackage[T1]{fontenc}

\usepackage[cmex10]{amsmath}  
\usepackage{amssymb}
\usepackage{subfigure,graphicx}

\newcommand{\E}{\mathbb{E}}

\newtheorem{Remark}{Remark}
\newtheorem{Theo}{Theorem}
\newtheorem{Prop}{Proposition}
\newtheorem{Lem}{Lemma}
\newtheorem{Cor}{Corollary}
\newtheorem{Def}{Definition}

\title{Robust Privatization with Multiple Tasks and the Optimal Privacy-Utility Tradeoff}
\author{Ta-Yuan Liu and I-Hsiang Wang,~\IEEEmembership{Member,~IEEE}
\thanks{This work was supported by NSTC of Taiwan under Grant 110-2634-F-002-029 and 111-2628-E-002-005-MY2 and NTU under Grant 112L893204. 
The material in this paper was presented in part at the 2020 IEEE Information Theory Workshop (ITW) \cite{liu2021privacy}.}
\thanks{T.-Y. Liu was with the Graduate Institute of Communication Engineering, National Taiwan University, Taipei, Taiwan. He is now with MediaTek Inc., Hsinchu, Taiwan (email: yorkrain@gmail.com).
}
\thanks{I.-H. Wang is with the Department of Electrical Engineering and the Graduate Institute of Communication Engineering, National Taiwan University, Taipei, Taiwan (email: ihwang@ntu.edu.tw).
}
}

\begin{document}
\allowdisplaybreaks
\maketitle

\begin{abstract}
In this work, fundamental limits and optimal mechanisms of privacy-preserving data release that aims to minimize the privacy leakage under utility constraints of a set of multiple tasks are investigated. 
While the private feature to be protected is typically determined and known by the sanitizer, the target task is usually unknown. 
To address the lack of information on the specific task, utility constraints laid on a set of multiple possible tasks are considered. The mechanism protects the specific privacy feature of the to-be-released data while satisfying utility constraints of all possible tasks in the set. First, the single-letter characterization of the rate-leakage-distortion region is derived, where the utility of each task is measured by a distortion function. It turns out that the minimum privacy leakage problem with log-loss distortion constraints and the unconstrained released rate is a non-convex optimization problem. Second, focusing on the case where the raw data consists of multiple independent components, we show that the above non-convex optimization problem can be decomposed into multiple parallel privacy funnel (PF) problems \cite{Makhdoumi2014Information} with different weightings. We explicitly derive the optimal solution to each PF problem when the private feature is a component-wise deterministic function of a data vector. The solution is characterized by a leakage-free threshold: when the utility constraint is below the threshold, the minimum leakage is zero; once the required utility level is above the threshold, the privacy leakage increases linearly. Finally, we show that the optimal weighting of each privacy funnel problem can be found by solving a linear program (LP).  A sufficient released rate to achieve the minimum leakage is also derived. Numerical results are shown to illustrate the robustness of our approach against the task non-specificity. 
\end{abstract}

\section{Introduction} \label{sec.into}
Data privacy has received great attention recently due to emerging applications of big data analytics. Owners of private data sets such as mobile users, schools, health care service providers, etc. are encouraged to release their data so that they can be utilized in certain tasks in data analysis, machine learning, etc.. However, this also increases the risk of privacy breach, and often the data need to be sanitized before the release for further utilization. 
As a result, there is an inevitable privacy-utility tradeoff. Characterizing the optimal tradeoff lies at the heart of research of data privacy. Among the vast literature, one of the most prominent privacy metric is \emph{differential privacy} \cite{Dwork2006calibrating}, which ensures the likelihood function of the released data remains roughly the same regardless of the realization (eg., the presence of a particular user's entry in a database) of the original one. Hence, from the privatized data release, it is difficult to make an inference about the raw data. Notably, due to the worst-case nature of this metric, the design of differentially private schemes does not require distributional information of the raw data. Such schemes are called \emph{prior-independent} \cite{HsuMartinez_21} and are known to suffer loss in utility \cite{friedman2010data,fredrikson2014privacy,farrand2020neither} and unsatisfactory privacy-utility tradeoff. The reason is that a prior-independent guarantee is laid on the privacy of the raw data itself, and such \emph{non-specific privacy} fundamentally reduces the utility of tasks that try to harness the data.

Meanwhile, in many scenarios, the sanitizer may aim to protect the privacy of specific features or covariates instead of the entire raw data. To achieve such \emph{specific privacy}, the design of privacy-preserving mechanisms may leverage the knowledge about the joint distribution of the private features and the raw data. 
In practice, the knowledge about the joint distribution could be incorporated from some training data set consisting of private and public covariates. The assumption of availability of such training data sets is well justified in various applications where the design of \emph{data-driven} privacy-preserving mechanisms is possible (to name a few, \cite{10.1007/978-3-030-01270-0_37,8919758,9578892}). 
Such \emph{prior-dependent} \cite{HsuMartinez_21} privacy-preserving mechanisms could potentially improve the privacy-utility tradeoff, especially when the private feature is not so informative for the target task. This direction has been taken in the information theory community, where information theoretic privacy metrics such as mutual information are adopted to quantify privacy leakage, that is, the amount of information that can be learned from the released data.   
In particular, in \cite{Sankar2013Utility}, the tradeoff between privacy and utility is studied through the lens of rate-distortion theory. The data is partitioned into public and private covariates, corresponding to utility and privacy respectively. In \cite{Makhdoumi2014Information}, the framework is further specialized to the case where mutual information is chosen as both the privacy and the utility metric, and the privacy funnel (PF) problem is formulated, where the aim is to minimize the amount of leakage in terms of mutual information between the private covariate and the released data subject to an utility constraint that is equivalent to taking logarithmic loss as the distortion function.

In most of the existing works regarding information theoretic privacy such as \cite{Sankar2013Utility,Makhdoumi2014Information}, privatized data release is tailored to meet a certain privacy leakage criterion while maximizing the utility of a \emph{specific task}, where the description of the task is given by specifying the distortion function for the reconstruction of certain covariates of the raw data, such as the logarithmic loss, the squared loss, etc.. 
As a result, the information theoretic privatization mechanisms derived from solving the corresponding optimization problems such as PF, depend on the specific task. In other words, there is an implicit assumption that the sanitizer is aware of the task in which the released data is utilized. 
This may not be realistic in many application scenarios such as collaborative data analytics from  privately-owned data sets because the downstream task may not be set at the time of privatization. In this case, it may be more desirable to develop privatization methods that achieve good privacy-utility tradeoff for a wide range of tasks. We term this paradigm ``\emph{non-specific task}.'' 

\subsection{Contributions}
The main focus of this work is on prior-dependent privacy-preserving data release mechanisms with specific privacy and non-specific tasks. 
Our goal is to develop robust privatization methods achieving the minimum privacy leakage while guaranteeing the utility of various tasks in a given set so that the data-sanitization mechanism is robust against the task non-specificity. 
Towards this goal, we employ an information theoretic framework extended from those in \cite{Sankar2013Utility,Makhdoumi2014Information}, with multiple target tasks and a single private feature to be protected. 
We first formulate a multi-letter information theoretic problem similar to that in \cite{Sankar2013Utility} and prove a single-letter characterization of the asymptotically optimal rate-leakage-distortion region. In the formulation of this multi-letter information theoretic problem, each ``letter'' is an ``entry'' of the data set, which can also be viewed as a symbol of a random source from the source coding perspective. Such a single-letter characterization gives the solid ground for the formulation of an optimization problem that aims to further characterize the optimal privacy-utility tradeoff when the data released rate is sufficiently large. This optimization problem can be viewed as a \emph{compound} privacy funnel \cite{Makhdoumi2014Information} problem. 
In general, the privacy funnel problem is a non-convex optimization problem, and the closed-form characterization of the privacy-utility tradeoff remains open \cite{Makhdoumi2014Information}, let alone the more complicated compound setting. 

To make progress and gain understanding in how the joint relationship among multiple target tasks impact the optimal privatization, we further restrict the investigation to the setting where each entry of the raw data set (that is, a single-letter random symbol representing the raw data) is a vector consisting of multiple independent components, serving as a canonical model for categorical data. 
Under this setting, when the private feature vector is a component-wise deterministic function of the raw data vector, we prove that a \emph{parallel} privatization mechanism that sanitizes each component independently is optimal, despite the fact that different possible tasks may be correlated. 
It is remarked that the proof of the optimality of parallel privatization is non-trivial even under the component-wise independence assumption. In the special case where the private feature vector to be protected is the entire raw data vector, this optimality can be proved in a straightforward manner by leveraging the chain rule of mutual information to construct for any given mechanism an equivalent parallel privatization scheme. In the more general case, however, direct extensions of the aforementioned argument fail, and the key is to identify the informative part in the raw data vector for multiple correlated tasks without further privacy leakage. Such a key step in our alternative construction is done by revoking the functional representation lemma (see \cite{1055989,1057042} and \cite[Appendix B]{Book_ElGamalNetwork}) for random variables. 

With the optimality of parallel privatization established, the original problem of finding the optimal privacy-utility tradeoff can be decomposed into multiple privacy funnel problems, each of which is associated with a different weighting. Each privacy funnel problem only involves a single component of the raw data vector, and its weighting corresponds to the amount of released information related to that component. A closed-form optimal solution to each parallel privacy funnel problem can then be derived explicitly, which is characterized intuitively by a ``leakage-free'' threshold explained as follows. 
When the utility requirement is below the threshold, there exists a zero-leakage privatization. When the utility requirement is above the threshold, the minimum privacy leakage is shown to be linearly proportional to the utility requirement. With the closed-form solution to each privacy funnel problem, it remains to find the optimal weightings, which can be done by solving a linear program (LP). A sufficient condition of the data released rate to achieve the optimal privacy-utility tradeoff is also derived. 
Numerical results are provided to illustrate the optimal privacy-utility tradeoff and demonstrate how the robustness is affected by the selection of the set of possible tasks. 

Compared to the conference version \cite{liu2021privacy}, this journal version expands in three aspects. First, in \cite{liu2021privacy}, the role of the data released rate in the optimal privacy-utility tradeoff problem is not addressed. In the journal version, we mend this gap by characterizing the optimal rate-leakage-distortion region and deriving the sufficient rate to achieve the optimal privacy-utility tradeoff. Second, we provide extension of the results partly to a setting with a stronger privacy guarantee. Finally, due to the space limit, \cite{liu2021privacy} only includes sketch proof of the main results, details of which are included in this version.

\subsection{Related works}
Apart from the works in data-driven privacy such as \cite{10.1007/978-3-030-01270-0_37,8919758,9578892} and those in the aforementioned information theoretic privacy \cite{Sankar2013Utility,Makhdoumi2014Information}, there have been some works in the literature considering protecting specific features rather than the entire raw data. 
In \cite{nageswaran2019data}, privacy and utility were measured in terms of the performance of certain statistical inference tasks: given a target probability of recovering a piece of information (the task) from the privatized data, the fundamental limit of the error probability of inferring the raw data at the adversary was derived. 
In \cite{geumlek2019profile}, the profile-based differential privacy is proposed to protect the identity of source distribution instead of data itself, and a higher utility is achieved by only obscuring those information related to the identity of the distribution. 
In \cite{lopuhaa2020privacy}, the adaption of differential privacy metric is considered, where the privacy leakage is measured with respect to the specific private feature rather than the raw data.

As for solving the privacy funnel problem, a greedy iterative algorithm by merging the output of privatizations in a pairwise manner was provided in \cite{Makhdoumi2014Information} without optimality guarantees. Such an algorithm was further improved by merging with an arbitrary number of combinations instead of pairwise ones at each iteration in \cite{ding2019submodularity}. To the best of our knowledge, efficient algorithms for solving the privacy funnel problem with optimality guarantees are still missing. The privacy funnel problem is also closely related to the information bottleneck problem \cite{tishby2000information}. 
In the original paper \cite{tishby2000information}, a convergent Blahut-Arimoto-type algorithm without optimality guarantees is provided to solve the problem. Recently, motivated by the global convergence results of alternating direction method of multipliers (ADMM) for non-convex objectives in some cases \cite{zhang2019fundamental}, ADMM is also applied to solve the information bottleneck problem \cite{bayat2019information, huang2021provably}. 
However, there is no guarantee for achieving the global optimum either. 

The information theoretic formulation can be viewed as a remote source coding problem \cite{DobrushinTsybakov_62,WolfZiv_70,Witsenhausen_80,Book_ElGamalNetwork} with multiple decoders and a privacy constraint, and hence is closely related to \cite{Kittichokechai2016Privacy}. 
The difference is that \cite{Kittichokechai2016Privacy} considered the single decoder model with side information and was focused on the minimum data released rate and the minimum distortion under a given privacy leakage constraint. Analytical results are mentioned only for some special examples such as binary and Gaussian models. In our work, multiple decoders are considered, and the focus is on minimizing the privacy leakage under given distortion constraints and sufficiently large released rate. Moreover, we derive analytical results for general distributions under a canonical framework of categorical data, which provides insight for the design of the privatization mechanism.

\subsection{Organization}
The rest of this paper is organized as follows. In Section \ref{sec.system model}, we introduce the system model and formulate the minimum leakage problem with the set of possible tasks. 
Section \ref{sec. main results} summarizes our main results and demonstrates the robustness of our proposed approach and the impact of the selection of the set of possible tasks. The detailed derivation and proofs of our main results that solve the minimum leakage problem are given in Section \ref{sec. parallelized} and \ref{sec. optimal privatizatoin}. In Section \ref{sec. parallelized}, we first show that the minimum leakage problem can be decomposed into multiple single-utility-constrained problems with different weightings. If the mutual-information-based privacy metric is considered, each of them can be regarded as a privacy funnel problem. Then, the solution to each privacy funnel problem and the optimal weighting is derived in Section \ref{sec. optimal privatizatoin}. An extended discussion of the sufficient condition of released rate to achieve the optimal privacy-utility tradeoff is given in Section \ref{sec. compress rate}. Finally, Section \ref{sec. conclusions} concludes the paper.

\section{Problem Formulation}\label{sec.system model}
In this section, let us first formulate a multi-letter information theoretic problem of privacy-preserving data release with multiple target tasks. The formulated problem, as commented in the previous section, is essentially a remote source coding problem with multiple reconstruction targets subject to an privacy leakage constraint. We then derive the single-letter characterization of the optimal rate-leakage-distortion region in the forthcoming Lemma~\ref{Lem. RLD region}, serving as the theoretical ground of the formulation of the optimization problem in \eqref{eq.mutual information opt problem} that solves optimal privacy-utility trade-off when specializing the distortion function as the logarithmic loss. While most derivations in this section are extensions of those in \cite{Sankar2013Utility,Makhdoumi2014Information,Kittichokechai2016Privacy}, we include them in the appendices for the sake of completeness.

\begin{figure*}[!t]
\centering
\includegraphics[width=0.75\linewidth]{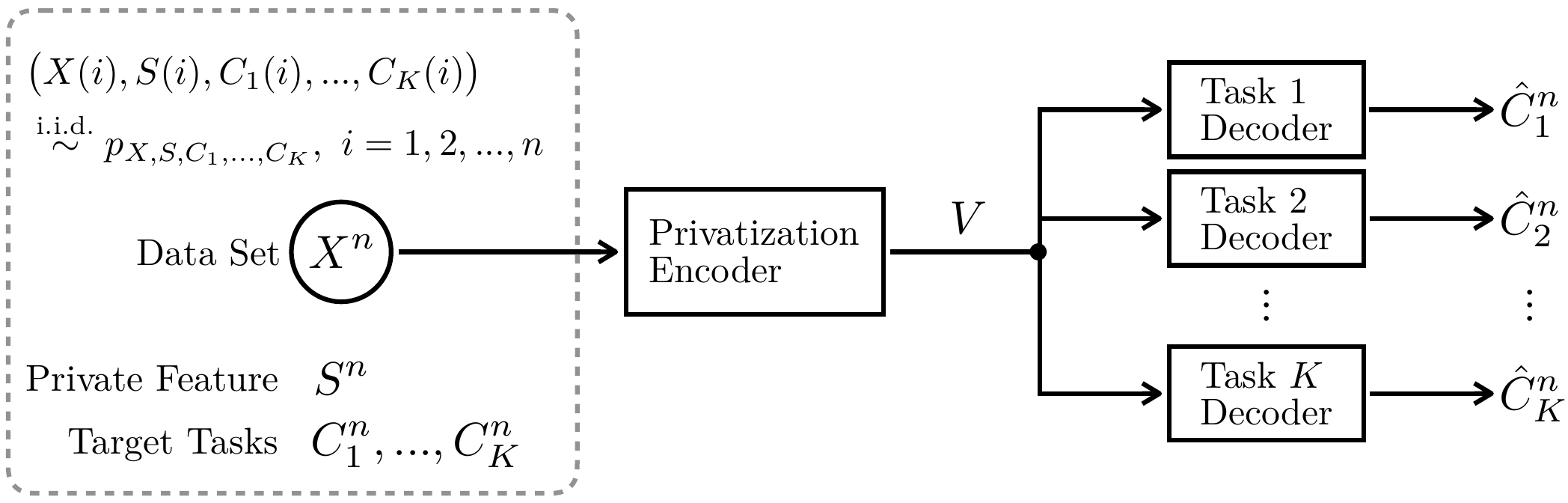}
\caption{The information theoretic multi-letter formulation of the privacy-preserving data release problem with $K$ possible tasks $\{C_1,...,C_K\}$ and a single private feature $S$.}
\label{fig: system model}
\end{figure*}

\subsection{An information theoretic formulation} \label{subsec. IT formulation}
The privacy-preserving data release system considered in this work is depicted in Figure~\ref{fig: system model}. In the system, a data set, viewed as a random sequence and denoted by $X^n =[(X(1),...,X(n)]$, is to be released for utilization. 
Prior to the release, the data set will be sanitized by a privatization encoder (mechanism) and become $V \in \mathcal{V}$, so that the private features $S^n = [S(1),...,S(n)]$ are leaked at an acceptable level, that is, the average mutual information per letter, $\frac{1}{n} I(S^n; V)$, is upper bounded by a prescribed threshold. In words, the privacy metric is chosen to be the mutual information. 
As for the utility, to address the non-specificity of the task in which the released data will be utilized, we take a \emph{compound} approach and introduce a set of $K$ possible target tasks, $\mathcal{T} = \{C_1,...,C_K\}$, and the goal of the $k$-th task, $k=1,2,...,K$, is to generate an estimate of the target sequence $C_k^n= [C_k(1),...,C_k(n)]$ from the released data $V$, denoted by $\hat{C}_k^n$, to within a certain distortion level. We assume that $X^n$, $S^n$, and $C_1^n,C_2^n,...,C_K^n$ are all length-$n$ discrete memoryless sequences, that is, i.i.d. across the $n$ letters, following a per-letter joint distribution $p_{X,S,C_1,...,C_K}(x,s,c_1,...,c_k)$. 
The alphabets of the data, the private feature, the target tasks, and the corresponding reconstructions are denoted by $\mathcal{X}$, $\mathcal{S}$, $\mathcal{C}_k$, and $\hat{\mathcal{C}}_k$, $k=1,...,K$, respectively. 
The utility metric of the $k$-th task is described by a per-letter distortion function $d_k: \mathcal{C}_k \times \hat{\mathcal{C}}_k \rightarrow [0, \infty)$, $k=1,...,K$. The distortion of a length-$n$ sequence is defined as the average of the per-letter distortion, following the usual convention in rate-distortion theory:
\[
      d_k^{(n)}(c_k^{n}, \hat c_k^{n}) \triangleq \frac{1}{n}  \sum_{i=1}^n d_k(c_{k}(i), \hat c_{k}(i)), ~~\forall k=1,...,K.
\]

The privatization mechanism (encoder) together with the $K$ reconstruction mechanisms (decoders) fully describe the scheme employed in the privacy-preserving data release system:
  \begin{Def}
    A $(|\mathcal{V}|, n)$-scheme for the privacy-preserving data release system consists of 
    \begin{enumerate}
    \item an encoder $\phi^{(n)}: \mathcal{X}^n\rightarrow \mathcal{V}$, and 
    \item $K$ decoders $\theta_k^{(n)}: \mathcal{V} \rightarrow \hat {\mathcal{C}}_k^{n} $, $k =1,...,K$.
    \end{enumerate}
  \end{Def}
The adopted information theoretic view is focused on the performance of the system in the asymptotic regime as $n\rightarrow \infty$. In the asymptotic regime, key parameters of concern are:
\begin{itemize}
\item Privacy leakage $L$.
\item Distortion levels $D_1,...,D_K$.
\item Released rate $R$. 
\end{itemize}

  \begin{Def}\label{def: rate leakage distortion}
    A rate-leakage-distortion tuple 
    \[
    (R,L,D_1,...D_K) \in \mathbb{R}^{K+2}_+
    \]  
    is said to be \emph{achievable} if for any $\delta > 0$ there exists a sequence of $(|\mathcal{V}|, n)$-schemes (indexed by $n$) such that for all sufficiently large $n$,
       \begin{align}
         &\frac{1}{n} \log |\mathcal{V}| \leq R + \delta, \label{eq. rate constraint} \\
         &\frac{1}{n} I(S^n;V) \leq L + \delta, \label{eq. leakage constraint}\\
         &\E[d_k^{(n)}(C_k^n, \theta_k^{(n)}(V))] \leq D_k +\delta, ~\forall k =1,...,K, \label{eq. distortion constraint} 
       \end{align}
       where $V = \phi^{(n)} (X^n)$. 
       The collection of all achievable rate-leakage-distortion tuple $(R,L,D_1,...D_K)$ is denoted as $\mathcal{R}$, the optimal rate-leakage-distortion region.
    \end{Def}

The main focus of this work is on the optimal tradeoff between privacy and utility. For this purpose, the minimum leakage subject to utility constraints, as defined in the following, will be of central interest in the rest of this paper.

\begin{Def}[Minimum privacy leakage]\label{def:privacy_leakage}
The minimum privacy leakage subject to utility constraints $(D_1,...,D_K)$ is defined as
\begin{equation}
L^*(D_1,...,D_K) \triangleq \inf_{(R,L)}\{ L \,\vert\, (R,L,D_1,...,D_K) \in \mathcal{R}\}.
\end{equation}
\end{Def}
Note that in the above definition, $L^*$ is not a function of the released rate $R$. Instead, it is the minimum leakage with arbitrarily large released rate. The impact of the released rate and the sufficient condition to achieve the optimal privacy-utility tradeoff will be studied in Section \ref{sec. compress rate}.

\subsection{A single-letter characterization of the optimal rate-leakage-distortion region}
The following lemma provides a single-letter characterization of the optimal rate-leakage-distortion region. 
While the single-letter characterization lays the theoretical ground towards further derivation of the optimal privacy-utility tradeoff, it is not the main contribution of this work since its proof is a straightforward extension of those single-letter characterizations in more simple settings in existing works \cite{Sankar2013Utility, Kittichokechai2016Privacy}. For the sake of completeness, we leave the details of the proof in Appendix~\ref{sec. proof of RLD region}.

\begin{Lem}\label{Lem. RLD region}
  The optimal rate-leakage-distortion region $\mathcal{R}$ is the collection of $(R,L,D_1,...D_K) \in \mathbb{R}^{K+2}_+ $ satisfying
  \begin{align} 
    & R \geq I(X;Y),  \label{eq. rate condition}\\
    & L \geq I(S;Y), \label{eq. leakage condition for region} \\
  & D_k \geq \E[d_k(C_k,\hat{C}_k)] , ~~\forall k=1,...,K, \label{eq. distortion condition for region}
  \end{align}
  for some $p_{Y,\hat{C}_1,...,\hat{C}_K|X}(y,\hat{c}_1,...,\hat{c}_k|x)$ where 
  \[
  (S,C_1,...,C_K) - X - Y - (\hat{C}_1,...,\hat{C}_K)
  \] 
  form a Markov chain, and $|\mathcal{Y}|\leq (|\mathcal{X}|+1) \cdot\prod_{k=1}^K |\hat{\mathcal{C}}_k|$.
\end{Lem}

\subsection{Specialization with the logarithmic loss}
With a general single-letter characterization of the optimal region in Lemma~\ref{Lem. RLD region}, in the following we further specialize it to the case where the distortion function is the logarithmic loss distortion. Log loss is widely used in learning theory \cite{Book_CesaBianchiPrediction} and source coding \cite{Sankar2013Utility,CourtadeWeissman_14, Kittichokechai2016Privacy,Book_ElGamalNetwork} and defined as follows:
  \begin{equation}
    d_k(c_k, \hat{c}_k) = \log \frac{1}{\hat{c}_k(c_k)}, \label{eq. def of log loss}
  \end{equation}
  where $\hat{c}_k(\cdot): \mathcal{C}_k \rightarrow [0,1]$ is a probability mass function over $\mathcal{C}_k$. 
Intuitively, the log-loss distortion qualifies a general ``soft'' estimate which provides the probability  instead of the deterministic value of the desired symbol.

With the log-loss distortion, the corresponding utility constraint in \eqref{eq. distortion condition for region} is shown to be equivalent to 
  \begin{align}
    I(C_k;Y) \geq \gamma(C_k) \triangleq H(C_k) - D_k, ~\forall k=1,...,K \label{eq. condition entropy constraint}
  \end{align}
since we can select the soft estimate $\hat{c}_k(c_k)$ as the conditional probability $p_{C_k|Y}(c_k|y)$. The detail of the proof can be found in Appendix \ref{sec. proof of equivalent of distortion constraint}. 
Accordingly, with a slight abuse of notation of $L^*$, the minimum privacy leakage defined in Definition~\ref{def:privacy_leakage} can be rewritten as 
  \begin{align}
    \notag    L^* = &\min_{p_{Y|X}}~ I(S;Y) \\
        &\mbox{~s.t.~}~ I(C_k;Y) \geq \gamma(C_k),  ~~\forall k = 1,...,K. \label{eq.mutual information opt problem}
  \end{align}
The optimization problem in \eqref{eq.mutual information opt problem} will be the main focus in the rest of the paper.

\subsection{Minimum leakage with component-wise data structure and specific privacy} 
\begin{figure*}[!t]
\centering
  \includegraphics[width=0.75\linewidth]{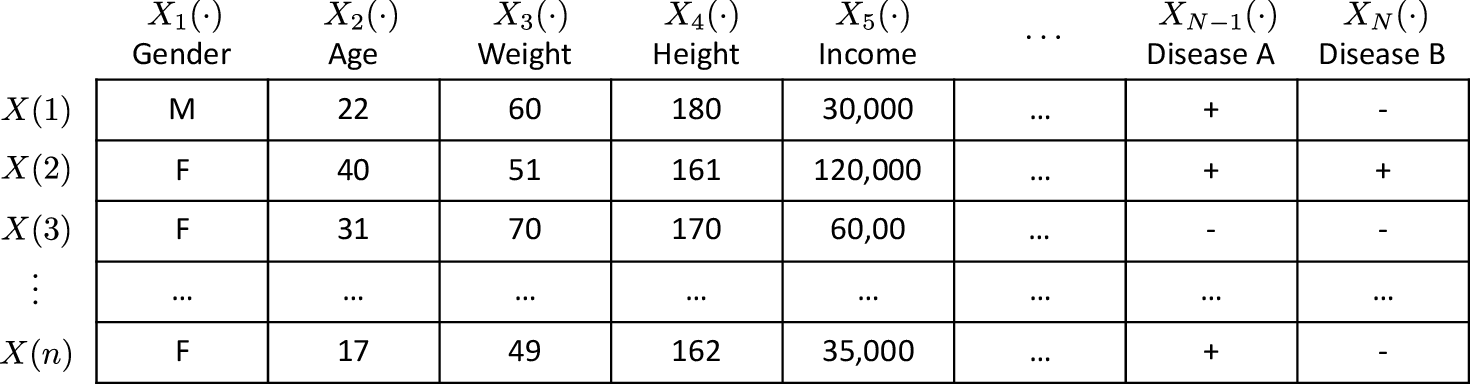}
  \caption{An example of a data set $X^n$. Each row represents a patient's entry, and different patients' entries are assumed to be i.i.d.. It is also assumed that each patient's entry consists of $N$ independent attributes. In short, the $i$-th row is the $i$-th patient's entry, and it is a vector $X(i)=[X_1(i),...,X_N(i)]$, $i=1,2,...,n$. 
The private feature $S$ in this example could be the disease(s) of the patient or the range of incomes (both are deterministic functions of $X$).}
\label{fig: Data Structure}
\end{figure*}

The problem in \eqref{eq.mutual information opt problem} aims to find the minimum leakage under multiple utility constraints $\gamma(C_1),...,\gamma(C_K)$, a compound version of the privacy funnel (PF) problem \cite{Makhdoumi2014Information}: 
when there is only one task $\mathcal{T} = \{X\}$, that is, the sole task is to reconstruct the entire raw data, the optimization problem in \eqref{eq.mutual information opt problem} degenerates to a PF. 
As mentioned in Section~\ref{sec.into}, PF is a non-convex optimization problem, and its closed-form solution remains open in general \cite{Makhdoumi2014Information}. 
To make progress, we further make the following assumptions:
  \begin{itemize}
  \item $X = [X_1, X_2, ..., X_N]$, where $X_1,...,X_N$ are mutually independent. 
  Note that here $X$ is a vector with $N$ components (coordinates), and it corresponds to a \emph{single letter} in the information theoretic formulation in Section~\ref{subsec. IT formulation}.
  \item $S = [S_1,S_2,...,S_N] = [f_1(X_1),f_2(X_2),...f_N(X_N)]$. 
  \item  $C_k$ is a sub-vector of $X$, $k=1,...,K$.
  \end{itemize}
In words, we assume that the raw data $X$ is a vector that consists of $N$ independent components (attributes), and the private feature $S$ is a component-wise deterministic function of $X$. An example is given in Fig. \ref{fig: Data Structure}. 
We also restrict to those tasks that can be represented by subsets of the $N$ components of the raw data vector $X$. 

\begin{Remark}
The restriction to the setup with the private feature vector being a component-wise function of the data vector is mainly for making progress in the open problem. It rule out the case where the private feature may be a function of more than one components. Nevertheless, our setting could provide stronger protection for such a situation. Let us explain it through a simple example where $X = [X_1,\ X_2,\ X_3]$ and $S = f(X_1, g(X_2), h(X_3))$. We may choose to protect more and set $S' = [X_1,\ g(X_2),\ h(X_3)]$. 
\end{Remark}

\begin{Remark}
The assumption of mutual independence of the $N$ components and the assumption that each of the possible tasks corresponds to the reconstruction of some components may not be valid in all applications. However, the setup serves as a vanilla model for the case where an entry of the data set can be decomposed into several independent parts and the tasks are correlated. As we will show later in the main results, when the data vector consists of independent attributes, we can decompose the our main problem in \eqref{eq.mutual information opt problem}, and thus, simplify the design of the privatization, even when the possible tasks are correlated with one another. 
\end{Remark}

\section{Main Results}\label{sec. main results}
In this section, we first summarize our main contributions in a series of theorems and corollaries, together with the ideas of the proofs. We then provide numerical results to illustrate the robustness of privatization against the task non-specificity as well as the impact of the set of possible tasks $\mathcal{T}$ on the privacy-utility tradeoff. Finally, the results are extended partly from the mutual information privacy to differential privacy.

\subsection{The optimal privacy-utility tradeoff}\label{subsec:opt_put}
The main contribution in this work is to solve the minimum leakage problem with multiple tasks described in \eqref{eq.mutual information opt problem} and obtain closed-form solutions that can bring insights into designing privatization that is robust with respect to the tasks. 
Our approach consists of several steps as summarized in the following theorems and corollaries. 

First, we show that a parallel privatization scheme $p_{Y|X}(y|x) = \Pi_{i=1}^{N}p_{Y_i|X_i}(y_i|x_i)$ is optimal for problem \eqref{eq.mutual information opt problem}.

\begin{Theo}\label{Theo.privated independently}
    For any feasible privatization $Y$ in \eqref{eq.mutual information opt problem}, there exists a parallel privatization $Y' = [Y'_1, Y'_2, ..., Y'_N]$ satisfying
    \begin{align}
      &p_{Y'|X}(y'|x) = \prod_{i=1}^{N} p_{Y'_i|X_i}(y'_i|x_i), \label{eq. parallized condtion 1}\\
      &I(S;Y') = I(S;Y),\label{eq. parallized condtion 2} \\
      &I(C_k;Y') \geq I(C_k;Y) , ~~~\forall k=1,...,K.\label{eq. parallized condtion 3}
    \end{align}
    \end{Theo} 
    
\begin{IEEEproof}[Sketch of Proof]
Let us sketch the idea of the proof and leave the details in Section \ref{subsec. proof of parallelized}. 
The idea is to first construct a $N$-component parallel $U=[U_1,...,U_N]$ which results in the same privacy leakage as $Y$. Then, by fixing the amount of leakage, we show that for each $(U_i, X_i)$, there exists a $Z_i$ which can release all the remaining amount of information of $X_i$, not included in $U_i$, without revealing additional privacy. As a result, the utility that $Y'=(U,Z)$ can achieve is not smaller than $Y$. 
\end{IEEEproof}
    
\begin{Remark}
It is noted that for the special case where $S=X$ and the possible tasks are all disjoint, that is, $C_k \cap C_k' = \emptyset$\footnote{For notational simplicity, here we view the sub-vector as a subset of the components of $X$.} $\forall\, k\neq k'$, the problem reduces to a source coding problem with multiple decoders, and independent encoding is optimal. 
We would like to stress that, however, the proof of the above key theorem (Theorem~\ref{Theo.privated independently}) is \emph{not} a straightforward extension of the aforementioned fact in source coding. 
To be more specific, for any non-parallel $p_{Y|X}$, one can easily construct a parallel version $Y'=[Y_1',...,Y_N']$ with $p_{Y'|X}(y'|x) = \prod_i p_{Y_i'|X_i}(y'_i|x_i)$ where $\mathcal{Y}'_i = \mathcal{X}_1 \times...\times \mathcal{X}_{i-1}\times \mathcal{Y}$ and
\[
\begin{aligned}
p_{Y'_i|X}(y'_i|x) &= p_{Y'_i|X}(x^{(i)}_1,...,x^{(i)}_{i-1},y^{(i)}|x_i)\\
&= p_{X_1,..,X_{i-1},Y|X_i}(x^{(i)}_1,...,x^{(i)}_{i-1},y^{(i)}|x_i),\ \forall\, i.
\end{aligned}
\]
Thus, by the independence of the components of $X$, we have 
\[
\begin{aligned}
I(X_i;Y_i') &= I(X_i;X_1,..,X_{i-1},Y)\\
&= I(X_i;Y|X_1,..,X_{i-1}),\ \forall\, i,
\end{aligned}
\]
and hence by the chain rule, $I(X;Y')=I(X;Y)$, and the leakage is maintained. Furthermore, for each task $C_k$, 
\[
\begin{aligned}
I(C_k; Y') &= \sum_{X_i\in C_k} I(X_i;Y'_i) \\
&\geq \sum_{X_i \in C_k} I(X_i; \{X_1,...,X_{i-1}\}\cap C_k, Y)\\
&= I(C_k; Y),
\end{aligned}
\]
and hence it satisfies all the utility constraints simultaneously. 
On the other hand, for the general case $S=f(X)$ and $C_k \cap C_k' \neq \emptyset$, this straightforward argument cannot be directly applied. Although the aforementioned parallel privatization $Y'$ satisfies all the utility constraints, it increases the privacy leakage: 
\[
\begin{aligned}
I(S;Y') &= \sum_i I(S_i;X_1,...,X_{i-1},Y)\\
 &\geq \sum_i I(S_i;S_1,...,S_{i-1},Y) = I(S;Y).
\end{aligned}
\] 
A simple alternative is to construct a parallel privatization by the similar argument with fixed privacy leakage, that is, construct parallel privatization $Y''=[Y_1'',...,Y_N'']$ where $\mathcal{Y}''_i = \mathcal{S}_1 \times...\times \mathcal{S}_{i-1}\times \mathcal{Y}$ and
\[ 
\begin{aligned}
p_{Y''_i|X_i}(y''_i|x_i) &= p_{Y''_i|X_i}(s^{(i)}_i,...,s^{(i)}_{i-1},y^{(i)}|x_i)\\
&= p_{S_1,...S_{i-1},Y|X_i}(s^{(i)}_i,...,s^{(i)}_{i-1},y^{(i)}|x_i),\ \forall\, i,
\end{aligned}
\] 
such that $I(S_i;Y_i'') = I(S_i;S_1,...,S_{i-1},Y)$, $\forall\, i$. This construction maintains the minimum privacy leakage. However, it is not guaranteed that it can satisfy the utility constraints, since $S_i$ may not carry the entire information of $X_i$ under the general setting $S= f(X)$. In short, our proof in the forthcoming Section \ref{subsec. proof of parallelized} can be regarded as a non-trivial extension of the argument in the special case where $S=X$.
\end{Remark}
 
Theorem~\ref{Theo.privated independently} asserts that optimal privatization can be generated separately with respect to each component, which also provides a clue of decomposing the optimization problem across the $N$ components $X_1,...,X_N$. 
Next, we leverage Theorem~\ref{Theo.privated independently} to further simplify the optimization problem. By introducing auxiliary variables $\alpha_1,...,\alpha_N$ and individual constraints on each component $I(X_i;Y_i) \geq \alpha_i$, $i=1,...,N$, the original minimum leakage problem can be rewritten as follows:
    \begin{align}
        \notag   L^*_A =&\min_{\{\alpha_i, p_{Y_i|X_i}\}_{i=1}^{N}}~ \sum_{i=1}^{N} I(S_i;Y_i) \\
            \notag &\mbox{~~~~~~~~~s.t.~}~~         \sum_{i:X_i \in C_k} \alpha_i \geq \gamma(C_k), ~~\forall k =1,...,K, \\
            \notag &~~~~~~~~~~~~~~~~~I(X_i;Y_i) \geq \alpha_i, ~~\forall i = 1,...,N, \\
            &~~~~~~~~~~~~~~~~~\alpha_i \geq 0 ~~\forall i = 1,...,N. \label{eq.parallel opt problem ver2}
        \end{align}
It leads to the following corollary regarding the equivalence between \eqref{eq.mutual information opt problem} and \eqref{eq.parallel opt problem ver2}.

\begin{Cor}\label{cor:equivalence}
The two optimization problems in \eqref{eq.mutual information opt problem} and \eqref{eq.parallel opt problem ver2} have the same minimum value, that is, 
\[
L^*_A = L^*.
\]
\end{Cor}        
\begin{IEEEproof}[Sketch of Proof]
Using the independence across components, one can decompose the leakage term into the sum of mutual information of all components and the utility term into the sum of corresponding individual constraints, respectively. The proof then follows straightforwardly. Details can be found in Section \ref{subsec. proof of parallelized Corollary}. 
\end{IEEEproof}
      
Corollary~\ref{cor:equivalence} suggests that once the values of the auxiliary $\{\alpha_i\}_{i=1}^N$ are given, the problem can be decomposed into $N$ parallel privacy funnel problems. 
The next step is to derive the solution to each privacy funnel problem. With our additional assumption that $S_i = f_i(X_i)$, $i=1,...,N$, the privacy funnel problem can be solved explicitly as shown in the following theorem.

\begin{Theo}\label{Theo.solution of privacy funnel}
Consider the privacy funnel problem defined as follows:
\begin{align}
  \notag    L_i^{PF}(\alpha_i) \triangleq &\min_{p_{Y_i|X_i}}~ I(S_i;Y_i) \\
    &\mbox{~s.t.~}~ I(X_i;Y_i) \geq \alpha_i. \label{eq.privacy funnel problem}
\end{align}
When $S_i = f_i(X_i)$, $i=1,...,N$, the minimum leakage is given as
\begin{align}
   L_i^{PF}(\alpha_i) =
  \begin{cases}
  0,&\text{if}~~ \alpha \leq \tau_i,\\
  \alpha_i - \tau_i,& \text{otherwise}.  \label{eq. minimum leakage of privacy funnel}
  \end{cases}
\end{align}
where $\tau_i = H(X_i) - H(S_i)$.
\end{Theo}

\begin{IEEEproof}[Sketch of Proof]
The proof of the theorem consists of two steps. We first show that it is possible to find a zero-leakage privatization $Y_i^f$ such that $Y_i^f$ and $S_i$ are mutually independent, and $I(X;Y_i^f)= \tau_i$. Then, we show that when $\alpha_i > \tau_i$, the minimum leakage can be achieved by releasing either $Y_i^f$ or the raw data $X_i$ in a randomized fashion. Details of the proofs can be found in Section \ref{subsec. achievable proof} and \ref{subsec. converse proof}, respectively. 
\end{IEEEproof}

For the privacy funnel problem in~\eqref{eq.privacy funnel problem}, the solution in \eqref{eq. minimum leakage of privacy funnel} is characterized by $\tau_i$, $i=1,...,N$, which we term the ``leakage-free'' threshold for each component. 
Zero leakage can be achieved if the individual constraint $\alpha_i$ is below the threshold, and the leakage increases linearly while the individual constraint is above the threshold. Under the assumption that $S_i = f(X_i)$, the leakage-free threshold $H(X_i)-H(S_i) = H(X_i|S_i)$ actually reflects the correlation between the raw data and the private feature. 
A higher correlation will result in a lower leakage-free threshold, which means that releasing data without privacy leakage is difficult, and vice versa. 

Finally, with Corollary~\ref{cor:equivalence} and Theorem~\ref{Theo.solution of privacy funnel}, the original minimum leakage problem in \eqref{eq.mutual information opt problem} can be simplified to a linear program as follows:

\begin{align}
\notag L^*_B = &\min_{\{\alpha_i\}_{i=1}^{N}}~ \sum_{i=1}^{N} \alpha_i - H(X_i) + H(S_i) \\
     \notag &\mbox{~~s.t.~}~ \sum_{i: X_i \in C_k}\alpha_i \geq \gamma(C_k), ~~\forall k=1,...,K,\\
     & ~~~~~~~~ \alpha_i \geq H(X_i) - H(S_i), ~~\forall i =1,...,N. \label{eq. allocation problem 2B}
  \end{align}
\begin{Cor}\label{cor:equivalence_2}
The two optimization problems in \eqref{eq.mutual information opt problem} and \eqref{eq. allocation problem 2B} have the same minimum value, that is, $L^*_B=L^*$.
\end{Cor}        
\begin{IEEEproof}[Sketch of Proof]
The intuition is that we can view $\alpha_i$ as the amount of released information related to $X_i$, and thus, there is no reason to choose $\alpha_i < \tau_i$ instead of $\alpha_i =\tau_i$, since both selections can achieve zero leakage. Hence, we can plug in $L^{PF}(\alpha_i) = \alpha_i -\tau_i$ and rewrite the problem as a linear program. Section \ref{subsec. optimal weighting} provides full details of the proof. 
\end{IEEEproof}   

\subsection{Robustness and the impact of the set of possible tasks}\label{sec. impact of possible set}
For numerical illustration, let us consider the scenario where the raw data consists of $5$ independent components, that is, $X=[X_1,...,X_5]$, and the private feature $S=[S_1,...,S_5]=[f_1(X_1),...,f_5(X_5)]$. 
For simplicity, the entropy of each component of the raw data and the private feature are assumed to be identical, that is $H(X_i) = 10$ and $H(S_i)=8$, $\forall\, i=1,...,5$. All the results in the following discussion are generated by applying the optimal privatization under the selected $\mathcal{T}$ and the utility constraint of those tasks in $\mathcal{T}$. For simplicity, we assume that utility constraints are proportional to the entropy of the task, i.e., $\gamma(C_k) = \gamma H(C_k)$, $\forall\, k$.

\begin{figure}[t]
    \begin{center}
      \includegraphics[width=\linewidth]{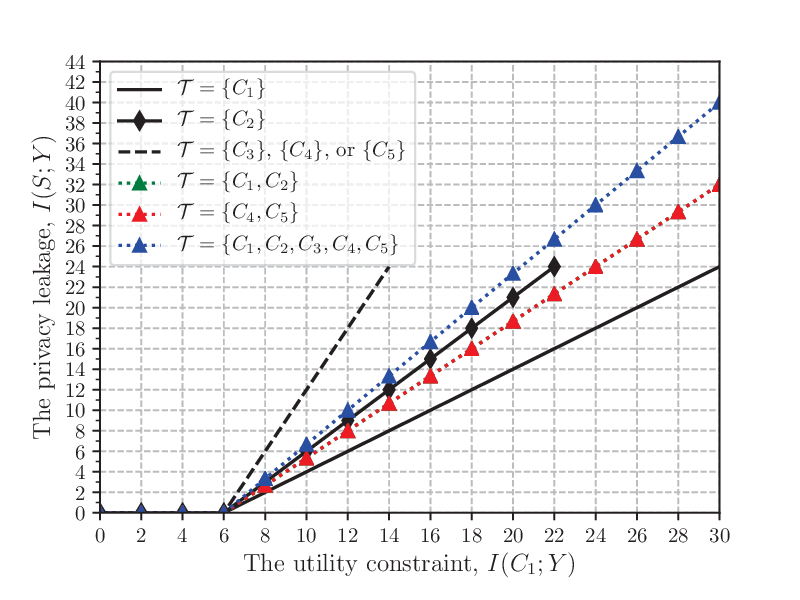}
      \caption{The privacy and utility tradeoff for the task $C_1$ under the privatization based on different possible sets.
       }\label{fig: tradeoff with different possible task}
    \end{center}
    \end{figure}

Fig. \ref{fig: tradeoff with different possible task} illustrates the privacy-utility tradeoff with different selections of the set of possible tasks, $\mathcal{T}$. We assume that there are $5$ candidates of the possible task considered by the sanitizer, that is, $C_1 = [X_1,X_2,X_3]$, $C_2=[X_2,X_3,X_4]$, $C_3= [X_1,X_4,X_5]$, $C_4=[X_2,X_4,X_5]$, and $C_5= [X_3,X_4,X_5]$. Let us further assume that the actual task is $C_1$. To achieve the same utility of the actual task, one needs to choose different $\gamma$'s, which result in different levels of privacy leakage, for different choices of the set of possible tasks $\mathcal{T}$.  
It can be shown that, in Fig. \ref{fig: tradeoff with different possible task}, one can always release $6$ information bits without any privacy leakage, regardless of the selection of $\mathcal{T}$.
Actually, this amount is equal to the leakage-free threshold of the true task, that is, $H(X_1,X_2,X_3)-H(S_1,S_2,S_3)$. Once the utility is beyond the threshold, zero privacy leakage is impossible, and the leakage increases linearly, with a slope that differs across different $\mathcal{T}$'s. 

Let us now illustrate the robustness of our approach compared to the approach that only considers a single specific task. For the single-task scenario, the sanitizer privatizes the data based on the selected task, and the performance depends on the correlation between the selected task and the actual task. 
The optimal tradeoff can be achieved with the correct selection $\mathcal{T} = \{C_1\}$, where the slope is equal to $1$. However, the performance drops quickly if the selection is wrong. The slope of the case where $\mathcal{T}=\{C_2\}$ is $3/2$. For the case with a lower correlated selection, e.g., $\mathcal{T}= \{C_3\}$, $\{C_4\}$, or $\{C_5\}$, where the selection covers only one component of $C_1$, the slope of the tradeoff is $3$, that is, one needs to leak $3$ bits of private information to increase only $1$ bit in utility. 

On the other hand, privatization with respect to multiple possible tasks is a more robust approach. It not only increases the chance to cover the actual task but also enhances the performance when the actual task is not included in the set. This can be seen by comparing the performance of $\mathcal{T}= \{C_4,C_5\}$ (including two choices with low correlation) and that of $\mathcal{T} =\{C_3\},\{C_4\}$ or $\{C_5\}$. 
Meanwhile, as the price to pay, considering multiple possible tasks will degrade the performance when the selection is precise, which can be observed by comparing the result of $\mathcal{T}= \{C_1\}$ and $\mathcal{T}= \{C_1,C_2\}$. For the most robust case $\mathcal{T}= \{C_1,...,C_5\}$, the information will be released from each component uniformly and the slope is $5/3$ no matter which one is the actual task.

\begin{figure}[t]
    \begin{center}
      \includegraphics[width=\linewidth]{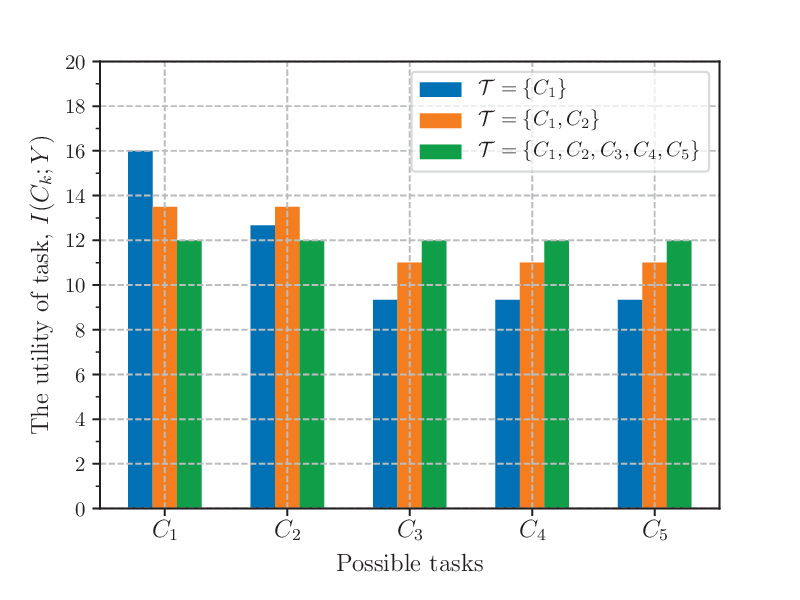}
      \caption{The effect of possible set of task on the utility of each task with fixed privacy leakage. 
       }\label{fig: the utility of each task under different possible sets}
    \end{center}
    \end{figure}
  
Fig. \ref{fig: the utility of each task under different possible sets} further illustrates the impact of $\mathcal{T}$ on the utility of each task. Under the constraint that the privacy leakage is $10$ bits, one can see that the privatization with respect to a single task, e.g., $\mathcal{T}=\{C_1\}$, is ``polarized'', in the sense that only the task that is strongly correlated with $C_1$ can achieve high utility. 
As more tasks are considered, e.g., $\mathcal{T}= \{C_1,C_2\}$, the performance of each task is relatively smooth. When the set $\mathcal{T}$ contains all the tasks, that is, $\mathcal{T}=\{C_1,...,C_5\}$, the amount of released information is uniform across all components, and each task has the same utility. 

\begin{figure}[t]
    \begin{center}
      \includegraphics[width=\linewidth]{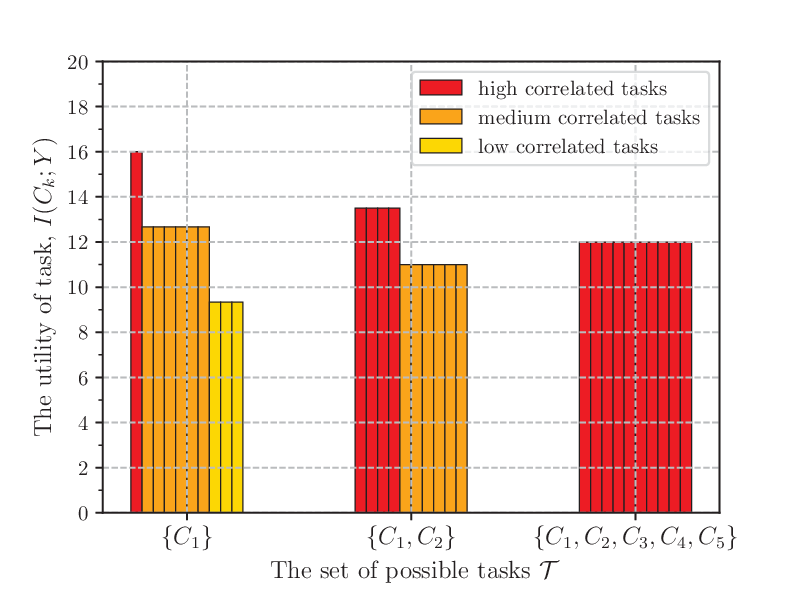}
      \caption{For given privacy leakage constraint, e.g., $10$, the utility of all the tasks containing $3$ components of $X$ with different choices of $\mathcal{T}$. 
       }\label{fig: probability of success versus diff Rc}
    \end{center}
    \end{figure}

Since the choice of $\mathcal{T}$ affects the utility of tasks, we can also qualify the choice of $\mathcal{T}$ based on the percentage of tasks which can be achieved. Fig. \ref{fig: probability of success versus diff Rc} illustrates this idea. 
In this figure, we consider all the tasks consisting of $3$ components of $X$ as possible candidates and assume each of them has an equal probability, that is, $1/10$, to be the actual task. Under a fixed privacy leakage constraint $10$, we can see that if the utility constraints are all below $12$, 
one should choose the largest possible set, that is, $\mathcal{T}=\{C_1,...,C_5\}$, such that the utility constraints of all possible tasks can be achieved. 
However, when the requirement increases, for example, $I(C;Y) = 13$, no task can be satisfied by such robust privatization based on $\mathcal{T}=\{C_1,...,C_5\}$. 
A better choice $\mathcal{T}=\{C_1,C_2\}$ provides a $4/10$ chance to achieve the utility of those tasks that are highly correlated with $C_1$ and $C_2$. 
When $I(C;Y) = 14$,  the utility constraint is achieved only if we can correctly select the actual task. 
A smaller one, for example, $\mathcal{T}= \{C_1\}$, should be considered, and it can meet the utility constraint with probability $1/10$.      

Let us close this subsection with a remark that summarizes the discussion on the robustness of our approach and the impact of the selection of the possible set $\mathcal{T}$.

\begin{Remark}
The privatization derived from solving the optimization problem \eqref{eq.mutual information opt problem} is a robust approach to achieve the requirement of an unknown task. Multiple utility constraints increase the chance to meet the requirement of the actual task even when the actual task is not included in the possible set. However, the additional consideration degrades the performance if the actual task can be  precisely known by the sanitizer beforehand. Hence, the choice of the set of possible tasks $\mathcal{T}$ depends on the tolerance of privacy leakage and the prior knowledge about the potential target tasks. Once a higher privacy leakage level can be tolerated, a larger set can be chosen such that it has a higher chance to cover the actual task. On the other hand, if the sanitizer has more knowledge about the actual task to be carried out, a smaller set can be chosen to render stronger privacy protection. 
\end{Remark}

\subsection{Extension beyond the mutual-information-based privacy}
The results in Section~\ref{subsec:opt_put} and \ref{sec. impact of possible set} are derived based on the mutual information privacy metric, which is an average-case leakage guarantee. Meanwhile, differential privacy (DP) provides a more strict leakage constraint with the following properties: (a) it protects the raw data instead of the specific privacy feature (b) it considers worst-case guarantees (c) it does not rely on the distribution of raw data, that is, it is prior-independent. 
However, as motivated in Section~\ref{sec.into}, in this work, we consider \emph{prior-dependent} schemes to enhance the privacy and utility tradeoff, and the focus is on the impact of the correlation among raw data, private feature, and tasks. Thus, following the prior-dependent paradigm, we make a slight extension of our results from the mutual-information-based framework to a framework called \emph{differential privacy with respect to $S$} (DP-S) \cite{lopuhaa2020privacy}. DP-S is an adaption of DP focused on the protection of a private feature instead of the raw data. The definition is given as follows.

\begin{Def}
     A privatization $p_{Y|X}(y|x)$ is said to achieve $\epsilon$-differential privacy with respect to $S$ if the induced conditional probability distribution $p_{Y|S}(y|s)$ satisfies
     \begin{align}
        p_{Y|S}(y|s) \leq p_{Y|S}(y|s') e^{( \epsilon \cdot {d_{H}(s,s')})},\quad \forall\, s,s',y,
     \end{align}
     where $d_{H}(s,s')$ denotes the Hamming distance between the two vectors $s$ and $s'$.
\end{Def}

Note that DP-S limits the ratio of $\sum_x p_{Y|X}(y|x)p_{X|S}(x|s)$ which implies the knowledge of $p_{X|S}$, and thus, it does not have the prior-independent property, even though $p_S$ can be arbitrary. Meanwhile, it still provides the worst-case guarantee over all possible realizations. The expression is slightly different from the standard form of DP, where the restriction of $s,s'$ being neighboring inputs is removed. Instead, in the leakage term, there is an extra Hamming distance $d_{H}(s,s')$ in the exponent. A kind of equivalence of this alternative definition and the original one was shown in \cite[Theorem 2.2]{dwork2014algorithmic}.

The privacy leakage (per different element) is now measured by $\epsilon$. Let us denote the minimum leakage (per different element) achieved by the privatization $p_{Y|X}(y|x)$ as 
\begin{align}
    \epsilon(p_{Y|X}) \triangleq \sup_{s,s',y: s\neq s'} \frac{1}{d_H(s,s')} \ln \frac{p_{Y|S}(y|s)}{p_{Y|S}(y|s')}.
\end{align}
Thus, the optimization problem for characterizing the optimal tradeoff is written as 
\begin{align}
 \notag   &\min_{p_{Y|X}} \epsilon(p_{Y|X})\\
& \mbox{~s.t.~~} I(C_k;Y) \geq \gamma(C_k), \forall k=1,...,K. \label{eq. DP tradeoff}
\end{align}
Let us extend the optimality of parallel privatization in Theorem \ref{Theo.privated independently}, as summarized in the proposition below.

\begin{Prop}\label{Cor. DP privated independently}
    For any feasible privatization $Y$ in \eqref{eq. DP tradeoff}, there exists a parallel privatization $Y' = [Y'_1, Y'_2, ..., Y'_N]$ satisfying
    \begin{align}
      &p_{Y'|X}(y'|x) = \prod_{i}^{N} p_{Y'_i|X_i}(y'_i|x_i), \label{eq. DP parallized condtion 1}\\
      &\epsilon(p_{Y'|X}) \leq \epsilon(p_{Y|X}),\label{eq. DP parallized condtion 2} \\
      &I(C_k;Y') \geq I(C_k;Y) , ~~~\forall k=1,...,K.\label{eq. DP parallized condtion 3}
    \end{align}
    \end{Prop}
    \begin{IEEEproof}[Sketch of Proof]
        This can be proved by a similar argument as that in the proof of Theorem \ref{Theo.privated independently}. Details can be found in Section \ref{subsec:pf_dp_parallel}.
    \end{IEEEproof}

Based on Proposition~\ref{Cor. DP privated independently}, we can similarly decompose the optimization problem in \eqref{eq. DP tradeoff} into multiple single-constraint problems: for $i=1,2,...,K$, 
\begin{align}
    \notag   &\min_{p_{Y_i|X_i}} \epsilon (p_{Y_i|X_i})\\
    & \mbox{~s.t.~~} I(X_i;Y) \geq \beta_i. \label{eq. DP tradeoff signle}
\end{align} 
However, in this case, deriving the closed-form solution to each single-constraint problem is challenging, and it remains an open problem. Instead, \cite{lopuhaa2020privacy} provides a numerical approach to solve the problem by finding the vertices of the feasible set which can be represented as a polytope. The original problem can be solved by searching the optimal weighting $\beta_i$ combining with the numerical approach given in \cite{lopuhaa2020privacy}.

\section{The Optimality of Parallel Privatization}\label{sec. parallelized}
One of the difficulties in deriving the optimal privatization mechanism of problem \eqref{eq.mutual information opt problem} comes from the dependency among multiple possible tasks. Imagine a simple case where all the possible tasks are disjoint, i.e., $C_k \cap C_{k'} = \emptyset$, $\forall\, k \neq k'$. Problem \eqref{eq.mutual information opt problem} can be easily split into to $K$ parts, due to the independence between $C_k$ and $C_{k'}$. Each part involves a single constraint $I(C_k;Y)\leq \gamma(C_k)$ and the corresponding object function $I(S_{C_k};Y)$, where $S_{C_k} = [S_i:X_i\in C_k]$ denotes the vector $S$ restricted onto $C_k$.  Then, the optimal privatization can be derived according to each single constraint problem, and it is a parallel mechanism.  
However, in practice, tasks are usually correlated and share common components of the raw data.  
In this section, we will show that even when the possible tasks are correlated, problem \eqref{eq.mutual information opt problem} can still be decomposed as multiple parallel problems, each of which has a single utility constraint. A key step is to establish Theorem~\ref{Theo.privated independently}, that is, to show that parallel privatization is optimal. We provide the proof in the following.

\subsection{Proof of Theorem \ref{Theo.privated independently}}\label{subsec. proof of parallelized}
For any feasible $p_{Y|X}(y|x)$, we prove that there exists $p_{Y'|X}(y'|x) =\prod_{i=1}^{N} p_{Y'_i|X_i}(y'_i|x_i)$ which satisfies all the utility constraints and achieves the same privacy leakage. We start the proof by constructing a $N$-component parallel $U=[U_1,...,U_N]$ which achieves the same privacy leakage, i.e., $I(S;U) = I(S;Y)$.
The alphabet of each component $U_i$ is given by $\mathcal{U}_i =\mathcal{Y} \times\mathcal{S}_1 \times ... \times\mathcal{S}_{i-1} $, $\forall i = 1,...,N$. The joint probability mass function is defined by
\begin{equation}
    p_{U, X, S, Y}(u,x,s,y) \triangleq p_{X,S,Y}(x,s,y)\prod_{i=1}^{N} p_{U_i|S_i}(u_i|s_i).  \label{parallelized joint pdf}
\end{equation}
Let $u_i = (s_1^{(i)},...,s_{i-1}^{(i)},y^{(i)})$. The conditional probability mass function $p_{U_i|S_i}(u_i|s_i)$ is defined as
\begin{align}
    &p_{U_i|S_i}(u_i|s_i)\notag\\
&\triangleq p_{S_1,...,S_{i-1},Y|S_i}(s_1^{(i)},...,s_{i-1}^{(i)},y^{(i)}|s_i),\ i=1,...,N. \label{conditional probabaility of Y given S}
\end{align}
From the from of the joint probability mass function \eqref{parallelized joint pdf}, we immediately have the following properties.
\begin{enumerate}
\item $U - S - (X,Y)$ form a Markov chain:
\par
\[
    \begin{aligned}
        &p_{U|X,S,Y}(u|x,s,y)\\ 
        &= \prod_{i=1}^{N} p_{U_i|S_i}(u_i|s_i) \\
        &= \frac{\sum_{x}\sum_{y} p_{X,S,Y}(x,s,y)\prod_{i=1}^{N}p_{U_i|S_i}(u_i|s_i)} {p_{S}(s)} \\
        &= \frac{\sum_{x}\sum_{y} p_{U,X,S,Y}(u,x,s,y)}{p_{S}(s)}\\
        &=  p_{U|S}(u|s).\label{eq. markov of S}
    \end{aligned}
\]
    
\item $\{(U_i,X_i,S_i)\}_{i=1}^{N}$ are mutually independent:
\par 
Due to the independence 
\[
p_{X,S}(x,s) = \prod_{i=1}^{N} p_{X_i,S_i}(x_i,s_i),
\] 
we have
    \begin{align}
\notag        &p_{U,X,S}(u,x,s)\\ 
\notag        &= p_{X,S}(x,s) \prod_{i=1}^{N} p_{U_i|S_i}(u_i|s_i)\\
\notag        &=\prod_{i=1}^{N} p_{X_i,S_i}(x_i,s_i) p_{U_i|S_i}(u_i|s_i)\\
        &=\prod_{i=1}^{N} p_{X_i,S_i}(x_i,s_i) p_{U_i|S_i, X_i}(u_i|s_i,x_i) \label{eq. Puxs} \\ 
        &=\prod_{i=1}^{N} p_{U_i,X_i,S_i}(u_i,x_i,s_i), \label{eq. independent of U}
    \end{align}
where the equality in \eqref{eq. Puxs} comes from $U - S - (X,Y)$.
Note that although in \eqref{conditional probabaility of Y given S}, we construct $U$ with the conditional probability mass function 
\[
p_{U_i|S_i}(u_i|s_i) = p_{S_1,...,S_{i-1},Y|S_i}(s_1^{(i)},...,s_{i-1}^{(i)},y|s_i),
\] 
it is not necessary to have the dependency between $U_i$ and $(S_1,...,S_{i-1})$. In fact, they are independent as shown in \eqref{eq. independent of U}. 
\end{enumerate}

According to \eqref{conditional probabaility of Y given S}, we have
\[
H(S_i|U_i) = H(S_i|S_1,...,S_{i-1},Y).
\] 
The privacy leakage can then be written as
\begin{align}
\notag    I(S;U) & = H(S) - H(S|U)\\
\notag    & = H(S) - \sum_{i=1}^{N} H(S_i|U_1,...U_N,S_1,...,S_{i-1})\\
    & =H(S) - \sum_{i=1}^{N} H(S_i|U_i) \label{eq. conditional entropy with mutual indep property}\\ 
\notag    & =H(S) - \sum_{i=1}^{N} H(S_i|S_1,...S_{i-1},Y)\\
\notag    & =H(S) - H(S|Y) = I(S;Y),
\end{align} 
where the equality in \eqref{eq. conditional entropy with mutual indep property} comes from \eqref{eq. independent of U}. 

To this end, we have constructed a parallel $U$ which contains the same amount of privacy as the given privatization $Y$. The next step is to show that we can find a parallel $Z=[Z_1,...,Z_N]$ which does not leak any further privacy, while $(U,Z)$ can achieve higher utility than $Y$. The construction of $Z$ relies on the following key lemma.

\begin{Lem}\label{Lem. existence of Z}
  For arbitrary jointly distributed discrete random variables $X$ and $Y$, there exists a discrete random variable $Z$ such that
  \begin{align}
    &H(X|Y,Z) = 0\quad \text{and} \label{eq. condition 1 of Z}\\
    &\text{$Z$ and $Y$ are independent}.   \label{eq. condition 2 of Z}
  \end{align}
\end{Lem}  

Note that Lemma~\ref{Lem. existence of Z} is a special case of the functional representation lemma \cite[Appendix B]{Book_ElGamalNetwork} and hence the proof is omitted. It asserts that for any random variables $X$ and $Y$, it is possible to extract all the remaining uncertainty of $X$ given $Y$ by an independent random variable $Z$. This means that the random variable $Z$ can release as much information of $X$ as possible and will not reveal anything about $Y$. 

Based on Lemma~\ref{Lem. existence of Z}, for each component $(U_i,X_i,S_i)$, we can find a random variable $Z_i$ such that
\begin{align}
    &H(X_i|U_i,S_i,Z_i) = 0 \label{eq. property of Z},\\
    &\text{$Z_i$ and $(U_i, S_i)$ are independent} \label{eq. property of Z 2}.
\end{align}
Since $\{(U_i, X_i, S_i)\}_{i=1}^{N}$ are independent, we can find $\{Z_i\}_{i=1}^{N}$ such that the property of independence still holds, that is, $\{(U_i, X_i, S_i,Z_i)\}_{i=1}^{N}$ are independent across different $i$'s.

We then construct the privatization $Y' = [Y'_1,...,Y'_N]$ where $Y'_i = (U_i, Z_i)$. $Y'$ is hence a ``parallel'' privatization, that is,
\[
\begin{aligned}
    p_{Y'|X}(y'|x) &= \prod_{i=1}^{N} p_{U_i,Z_i|X_i}(u_i,z_i|x_i)\\
    &= \prod_{i=1}^{N} p_{Y'_i|X_i}(y'_i|x_i).
\end{aligned}
\]
Also, we can show that both privacy and utility requirements \eqref{eq. parallized condtion 2} and \eqref{eq. parallized condtion 3} hold as follows. 
Since $Z_i$ is independent of $(S_i, U_i)$, it does increase any privacy leakage: 
\begin{align}
\notag    I(S;Y') & = \sum_{i=1}^{N} I(S_i;U_i,Z_i)
 = \sum_{i=1}^{N} I(S_i;U_i)\\
    & = I(S;U) = I(S;Y). \label{eq. parallelized privacy constraint}
\end{align}
Thus, $Y'$ contains the same amount of privacy as $Y$. Moreover, due to the fact that all information related to privacy is released through $Z$, we can show that $Y'$ provides at least the same level of utility as $Y$ does:
\begin{align}
\notag    & I(C_k;Y')\\ 
\notag    & =  H(C_k) - H(C_k|Y')\\
\notag    & = H(C_k) - \sum_{i:X_i\in C_k}  H(X_i |U_i,Z_i)\\
    & = H(C_k) - \sum_{i:X_i\in C_k}  H(X_i, S_i|U_i,Z_i) \label{eq. parallelized utility constraint 1}\\
\notag    & = H(C_k) - \sum_{i:X_i\in C_k}  H(X_i|U_i,Z_i, S_i) + H(S_i|U_i,Z_i)\\
    & = H(C_k) - \sum_{i:X_i\in C_k}  H(S_i|U_i,Z_i)\label{eq. parallelized utility constraint 2}\\
\notag    & = H(C_k) - \sum_{i:X_i\in C_k}  H(S_i|U_i)\\
\notag    & = H(C_k) - \sum_{i:X_i\in C_k}  H(S_i|S_1,...,S_{i-1},Y)\\
\notag    & \geq H(C_k) -\sum_{i:X_i\in C_k}  H(S_i|\{S_1,...,S_{i-1}\}\cap S_{C_k},Y)\\
\notag    & = H(C_k) - H(S_{C_k}|Y)\\
    & \geq H(C_k) - H(C_k|Y)\label{eq. parallelized utility constraint 3}\\
    & = I(C_k;Y), \label{eq. parallelized utility constraint 4}
\end{align}
where $S_{C_k}$ denotes $[S_i:X_i\in C_k]$. Equation \eqref{eq. parallelized utility constraint 1} and \eqref{eq. parallelized utility constraint 3} follow from the fact $S_i = f_i(X_i)$ and \eqref{eq. parallelized utility constraint 2} follows from \eqref{eq. property of Z}. Combining \eqref{eq. parallelized privacy constraint} and \eqref{eq. parallelized utility constraint 4}, we show that $Y'$ is the desired parallel privatization and complete the proof.

\subsection{Proof of Corollary \ref{cor:equivalence}}\label{subsec. proof of parallelized Corollary}
Based on Theorem \ref{Theo.privated independently}, the optimization problem in \eqref{eq.mutual information opt problem} can be rewritten as the following problem
\begin{align}
\notag    L^*=&\min_{\{p_{Y_i|X_i}\}_{i=1}^{N}}~ \sum_{i=1}^{N} I(S_i;Y_i) \\
    &\mbox{~~~~~~~s.t.~}~ \sum_{i:X_i \in C_k}  I(X_i;Y_i) \geq \gamma(C_k),\ k = 1,...,K.  \label{eq.parallel opt problem}
\end{align}
By introducing the individual constraint on each component $I(X_i;Y_i) \geq \alpha_i$, we can replace the utility constraints by $\sum_{i:X_i\in C_k} \alpha_i \geq \gamma(C_k),~ \forall k =1,...,K$. This also proves the equivalence between \eqref{eq.parallel opt problem} and \eqref{eq.parallel opt problem ver2} and thus, $L^* = L^*_A$, which completes the proof of Corollary \ref{cor:equivalence}.

$\{\alpha_i\}_{i=1}^N$ can be regarded as the weightings of the $N$ single-constraint problems to be written, each of which represents the amount of the released information that is related to the corresponding component $X_i$, $=1,2,...,N$. Once the weightings $\{\alpha_i\}_{i=1}^N$ are given, the optimization problem can be decomposed into $N$ parallel single-constraint problems. Each of them is actually the privacy funnel problem as given below:
\begin{align}
  \notag    \min_{p_{Y_i|X_i}}~ &I(S_i;Y_i) \\
  \notag    \mbox{~~s.t.~~}~ &I(X_i;Y_i) \geq \alpha_i.  
\end{align}
Thus, towards solving the minimum-leakage problem, the remainder can be split into two phases: 
one is to solve each of the parallel single-utility-constraint problems (that is, parallel PFs), and the other is to find the optimal weightings $\{\alpha_i\}_{i=1}^{N}$. 
Both of them will be thoroughly explored in Section~\ref{sec. optimal privatizatoin}.

\subsection{Proof of Proposition \ref{Cor. DP privated independently}}\label{subsec:pf_dp_parallel}
We prove this proposition with the similar steps as the proof of Theorem \ref{Theo.privated independently}. For any feasible $p_{Y|X}(y|x)$, we can construct $U = [U_1,...,U_N]$ by \eqref{parallelized joint pdf} and \eqref{conditional probabaility of Y given S}. Let the desired parallel privatization $Y'=[Y'_1,...,Y'_N]$ where $Y'_i = (U_i, Z_i)$ and $\{Z_i\}_{i=1}^N$ are the independent random variables with properties \eqref{eq. property of Z} and \eqref{eq. property of Z 2}. As shown in the proof of Theorem \ref{Theo.privated independently}, we know that $Y'$ satisfies \eqref{eq. DP parallized condtion 1} and \eqref{eq. DP parallized condtion 3}. The rest of proof is to show \eqref{eq. DP parallized condtion 2}, that is, $\epsilon(p_{Y'|X}) \leq \epsilon(p_{Y|X})$.

For any realization $y',s, s'$, we have 
\begin{equation}
    \frac{p_{Y'|S}(y'|s)}{p_{Y'|S}(y'|s')} = \frac{\frac{p_{S|Y'}(s|y')}{p_{S}(s)}}{\frac{p_{S|Y'}(s'|y')}{p_{S}(s')}}
    \overset{(a)}{=} \frac{\frac{p_{S|U}(s|u)}{p_{S}(s)}}{\frac{p_{S|U}(s'|u)}{p_{S}(s')}}
    = \prod_{i=1}^N\frac{\frac{ p_{S_i|U_i}(s_i|u_i)}{ p_{S_i}(s_i)}}{\frac{ p_{S_i|U_i}(s_i'|u_i)}{ p_{S_i}(s_i')}}. \label{eq. DP proof 1}
\end{equation}
Here $(a)$ follows from $Y' = (U,Z)$ and the fact that $Z$ and $(S, U)$ are independent. Let $u_i = (s_1^{(i)},...,s_{i-1}^{(i)},y^{(i)})$. 

\begin{figure*}[!t]
\normalsize
\begin{align}
\notag    \frac{\frac{p_{S_i|U_i}(s_i|u_i)}{ p_{S_i}(s_i)}}{\frac{p_{S_i|U_i}(s_i'|u_i)}{p_{S_i}(s_i')}}
    & =\frac{\frac{ p_{S_i|S_1,...,S_{i-1},Y}(s_i|s_1^{(i)},...,s_{i-1}^{(i)},y^{(i)})}{ p_{S_i}(s_i)}}{\frac{ p_{S_i|S_1,...,S_{i-1},Y}(s_i'|s_1^{(i)},...,s_{i-1}^{(i)},y^{(i)})}{ p_{S_i}(s_i')}}
 =\frac{
    \frac{p_{S_i|S_1,...,S_{i-1},Y}(s_i|s_1^{(i)},...,s_{i-1}^{(i)},y^{(i)})}
    {p_{S_i}(s_i)}
    \frac{p_{S_1,...,S_{i-1}|Y}(s_1^{(i)},...,s_{i-1}^{(i)})}
    {p_{S_1,...,S_{i-1}}(s_1^{(i)},...,s_{i-1}^{(i)})}}
     {\frac{ p_{S_i|S_1,...,S_{i-1},Y}(s_i'|s_1^{(i)},...,s_{i-1}^{(i)},y^{(i)})}{ p_{S_i}(s_i')}\frac{p_{S_1,...,S_{i-1}|Y}(s_1^{(i)},...,s_{i-1}^{(i)})}
     {p_{S_1,...,S_{i-1}}(s_1^{(i)},...,s_{i-1}^{(i)})}}\\
\notag     & =\frac{ p_{Y|S_1,...,S_{i-1},S_i}(y^{(i)}|s_1^{(i)},...,s_{i-1}^{(i)},s_i)}
      {  p_{Y|S_1,...,S_{i-1},S_i}(y^{(i)}|s_1^{(i)},...,s_{i-1}^{(i)},s_i')}    \\
\notag      & =\frac{ \sum_{s_i,...,s_N} p_{Y|S_1,...,S_N}(y^{(i)}|s_1^{(i)},...,s_{i-1}^{(i)},s_i,...,s_N) p_{S_{i+1},...,S_N}(s_{i+1},...,s_N)}
      { \sum_{s_i,...,s_N} p_{Y|S_1,...,S_N}(y^{(i)}|s_1^{(i)},...,s_{i-1}^{(i)},s'_i,...,s_N) p_{S_{i+1},...,S_N}(s_{i+1},...,s_N)}    \\
      &\leq  e^{\epsilon(p_{Y|X}) \cdot d_H(s_i,s'_i)}. \label{eq. DP proof 2}
\end{align} 
\hrulefill
\vspace*{4pt}
\end{figure*}

We rewrite the $i$-th term of \eqref{eq. DP proof 1} as shown on the top of the next page. 
The inequality in \eqref{eq. DP proof 2} comes from the fact 
\[
p_{Y|S}(y|s) \leq p_{Y|S}(y|s') e^{\epsilon(p_{Y|X}) \cdot d_H(s,s')}.
\] 
Combining \eqref{eq. DP proof 1} and \eqref{eq. DP proof 2}, we have $\forall\, s,s',y'$, 
\[
    \frac{p_{Y'|S}(y'|s)}{p_{Y'|S}(y'|s')} \leq \prod_i e^{\epsilon(p_{Y|X}) \cdot d_H(s_i,s'_i)}
    = e^{\epsilon(p_{Y|X})\cdot d_H(s,s')}, 
\]
which implies $\epsilon(p_{Y'|X}) \leq \epsilon(p_{Y|X})$ and completes the proof.

\section{An Explicit Solution to the Privacy Funnel and the Optimal Weightings} \label{sec. optimal privatizatoin}
In this section, we derive the optimal solution to each of the parallel single-utility-constraint privacy funnel problems and prove Theorem~\ref{Theo.solution of privacy funnel}. With the assumption $S_i= f_i(X_i)$, $i=1,2,...,N$, each privacy funnel problem can be solved explicitly. We then prove Corollary~\ref{cor:equivalence_2} and show that the minimum leakage problem is equivalent to a linear program.

\subsection{Achievability proof of Theorem \ref{Theo.solution of privacy funnel}}\label{subsec. achievable proof}

Let us consider two different cases (i) $\alpha_i \leq H(X_i) - H(S_i)$ and  (ii) $\alpha_i > H(X_i) - H(S_i)$, respectively. We first provide a leakage-free privatization for the first case. Then, we show that the optimal privatization for the second case can be generated by releasing the leakage-free privatization and the raw data in a randomized fashion.  

\subsubsection{$\alpha_i \leq H(X_i)- H(S_i)$}
The construction of leakage-free privatization directly follows from the results in Lemma \ref{Lem. existence of Z}. For any $(X_i, S_i)$, we can find $Y_i^{f}$ independent of $S_i$ such that $H(X_i|S_i,Y_i^{f}) = 0$. Thus, we have $I(S_i;Y_i^{f}) =0$ and
\begin{align*}
    I(X_i;Y_i^{f}) &= H(X_i) - H(X_i|Y_i^{f})\\
           &= H(X_i) - H(X_i,S_i|Y_i^{f})\\
           &= H(X_i) - H(X_i|S_i, Y_i^{f})- H(S_i|Y_i^{f})\\
           &= H(X_i) - H(S_i|Y_i^{f})\\
           & =H(X_i) - H(S_i) \geq \alpha_i,
\end{align*} 
which shows that  $Y_i^{f}$ is the desired leakage-free privatization.

\subsubsection{$\alpha_i > H(X_i)-H(S_i)$}
In this case, we show that the minimum leakage can be achieved by alternately releasing $Y_i^{f}$ and $X_i$ with a certain probability. Let $Y'_i$ be the privatization and 
\[
   Y'_i = 
\begin{cases}
Y_i^{f} ,&\text{with prob.}~~ p,\\
X_i , & \text{with prob.}~~1-p,  
\end{cases}
\]
with $p = (H(X_i)-\alpha_i) / H(S_i)$. Clearly, we have
\[
  I(X_i;Y'_i) = pI(X;Y_i^{f}) + (1-p) H(X_i)= \alpha_i,
\]
and
\begin{align*}
  I(S_i;Y'_i) &= pI(S_i;Y_i^{f}) + (1-p) H(S_i)\\
  & = \alpha_i - H(X_i) + H(S_i),
\end{align*}
which completes the proof.

In summary, we show that there exists a leakage-free privatization $Y^{f}_i$ which can release at most $H(X_i)-H(S_i)$ bits of information. When the utility constraint is larger than the leakage-free threshold, randomly releasing raw data $X_i$ and the leakage-free privatization $Y^{f}_i$ with proper probabilities achieves a linear privacy-utility tradeoff which can be shown to be optimal by combining the following converse.

\subsection{Converse proof of Theorem \ref{Theo.solution of privacy funnel}}\label{subsec. converse proof}
The converse is straightforward. Note that $I(S_i;Y_i) \geq 0$ and
\begin{align}
\notag   I(S_i;Y_i) &= H(S_i) - H(S_i|Y_i)\\
   & \geq H(S_i) - H(X_i|Y_i) \label{eq. converse proof deterministic f}\\
\notag   & = I(X_i;Y_i) - H(X_i) + H(S_i) \\
\notag   & \geq \alpha_i - (H(X_i) - H(S_i)),
 \end{align}
where \eqref{eq. converse proof deterministic f} comes from $S_i=f(X_i)$. Thus, \eqref{eq. minimum leakage of privacy funnel} is a lower bound of the minimum privacy leakage.

\subsection{Proof of Corollary~\ref{cor:equivalence_2}}\label{subsec. optimal weighting}
With Theorem \ref{Theo.solution of privacy funnel}, the optimization problem \eqref{eq.parallel opt problem ver2} can be rewritten as the following ``allocation'' problem
\begin{align*}
&\min_{\{\alpha_i\}_{i=1}^{N}}~ \sum_{i=1}^{N} L_i^{PF}(\alpha_i) \\
&\mbox{~~s.t.~}~ \sum_{i: X_i \in C_k}\alpha_i \geq \gamma(C_k), ~~\forall k=1,...,K,\\
& ~~~~~~~~ \alpha_i \geq 0, ~~\forall i =1,...,N. 
\end{align*}

What remains is to derive the optimal allocation $\{\alpha_i\}_{i=1}^{N}$. The variable $\alpha_i$ and $L_i^{PF}(\alpha_i)$ represent the amount of released information and the privacy leakage via $Y_i$, respectively. Towards achieving the optimal privacy-utility tradeoff, there is no reason to select $\alpha_i$ less than the ``leakage-free'' threshold $\tau_i = H(X_i) - H(S_i)$. Hence, we have 
\[
L_i^{LP}(\alpha_i) = \alpha_i - H(X_i) + H(S_i),
\] 
and the problem is turned into a linear programing (LP) problem:
\begin{align}
   \notag  &\min_{\{\alpha_i\}_{i=1}^{N}}~ \sum_{i=1}^{N} \alpha_i - H(X_i) + H(S_i) \\
     \notag &\mbox{~~s.t.~}~ \sum_{i: X_i \in C_k}\alpha_i \geq \gamma(C_k), ~~\forall k=1,...,K,\\
     & ~~~~~~~~ \alpha_i \geq H(X_i) - H(S_i), ~~\forall i =1,...,N. \label{eq. allocation problem 2}
  \end{align}
It can be observed from \eqref{eq. allocation problem 2} that the privacy-utility tradeoff for each component is the same and the tradeoff ratio is $1$ once the amount of released information is larger than the leakage-free threshold. However, releasing information via those components included by more tasks in the set $\mathcal{T}$ can contribute to meet more utility constraints. This suggests that one should release more information from those components included by more tasks in $\mathcal{T}$. Unfortunately, even with this intuition, the closed-form solution to the linear program is difficult to derive since it depends on the set $\mathcal{T}=\{C_1,..,C_K\}$ in a case-by-case manner. Nevertheless, one can efficiently solve this LP problem by standard techniques such as the simplex algorithm.

\section{Sufficiency of the Released Rate} \label{sec. compress rate}
In addition to the privacy and utility, the released rate, which represents the necessary number of bits per letter to transmit the privatized data, is also an important metric of a privatization mechanism as mentioned in Section \ref{subsec. IT formulation}. We have studied the optimal tradeoff between privacy and utility. The next interesting problem is to know what is the minimum released rate to achieve the optimal privacy-utility tradeoff. 

To address this problem, let us define the minimum released rate under the given utility constraints $\bar \gamma \triangleq [\gamma(C_1),...,\gamma(C_K)]$ and the corresponding minimum privacy $L^*(\bar \gamma)$ (the minimum value of problem \eqref{eq.mutual information opt problem}):
\begin{align*}
  R(\bar{\gamma}, L^*(\bar \gamma)) \triangleq& \min_{p_{Y|X}} I(X;Y)\\
  & \mbox{s.t.}~~ I(C_k;Y) \geq \gamma(C_k), \forall k=1,...,K, \\
  & ~~~~~~ I(S;Y) \leq L^*(\bar \gamma). 
\end{align*}
Unfortunately, to derive an analytical form of $R(\bar{\gamma}, L^*(\bar \gamma))$ is a challenge as we fail to show that parallel privatization is optimal for the above minimum released rate problem. Also, the lack of closed-form solutions to $L^*(\bar \gamma)$, which can only be derived by solving a linear problem as shown in the previous section, makes this problem intractable. To make progress, we consider the minimum released rate with an additional constraint on each component, that is,
\begin{align}
  \notag R'(\bar{\gamma}, L^*(\bar \gamma)) \triangleq& \min_{p_{Y|X}} I(X;Y)\\
  \notag & \mbox{s.t.}~~ I(C_k;Y) \geq \gamma(C_k), \forall k=1,...,K, \\
  \notag  & ~~~~~~  I(X_i;Y_i) \geq \tau_i, \forall i=1,...,N, \\
  & ~~~~~~ I(S;Y) \leq L^*(\bar \gamma), \label{eq. problem of compression rate with additional constraints}
\end{align}
where $\tau_i = H(X_i)-H(S_i)$ is the leakage-free threshold of component $i$, $i=1,2,...,N$. Note that the additional constraint will not increase the privacy leakage. This is because one can always release $\tau_i$ amount of information about component $i$ without any privacy leakage. Problem \eqref{eq. problem of compression rate with additional constraints} considers a setting where all the information which is not related to the private feature is required to be released. This enhances the robustness of the privatization mechanism. For example, consider an extreme case where the actual task and the set of possible tasks have no overlap. With these additional constraints, at least part of the data which does not include the private feature will be released. As the price to pay, a higher released rate is necessary to satisfy those additional constraints. 

The more stringent problem \eqref{eq. problem of compression rate with additional constraints} can be solved by the following theorem.
\begin{Theo}\label{Lem. min privatization rate}
There exist a parallel privatization 
\[
p_{Y|X}(y|x) =\Pi_{i=1}^N p_{Y_i|X_i}(y_i|x_i)
\] 
that can achieve the minimum released rate in problem \eqref{eq. problem of compression rate with additional constraints}, and the minimum rate is 
\[
    R'(\bar{\gamma}, L^*(\bar \gamma)) =L^*(\bar \gamma) + \sum_{i=1}^N \tau_i.
\]
\end{Theo}
\begin{IEEEproof}
  We prove this theorem by considering two minimum rate problems which can be regarded as the upper and lower bound of $R_{p}(\bar{\gamma}, L^*(\bar \gamma))$, respectively.  Both minimum rate problems are under the restriction of using parallel privatization. 
  
The first one is
  \begin{align}
    \notag  R'_{p}(\bar{\gamma}, L^*(\bar\gamma)) =& \min_{p_{Y|X} =\prod_i p_{Y_i|X_i}} I(X;Y)\\
    \notag & \mbox{s.t.}~~ I(C_k;Y) \geq \gamma(C_k), \forall k=1,...,K, \\
    \notag & ~~~~~~  I(X_i;Y_i) \geq \tau_i, \forall i=1,...,N, \\
    & ~~~~~~ I(S;Y) \leq L^*(\bar{\gamma}), \label{eq. compression with parallelized}
  \end{align} 
  where the subscript $p$ stands for the restriction of using parallel privatization.
  The second one considers the case without privacy constraint, that is, 
  \begin{align}
    \notag R_{p}'(\bar{\gamma}, H(S)) = & \min_{p_{Y|X} =\prod_i p_{Y_i|X_i}} I(X;Y)\\
    \notag& \mbox{s.t.}~~ I(C_k;Y) \geq \gamma(C_k), \forall k=1,...,K, \\
      & ~~~~~~  I(X_i;Y_i) \geq \tau_i, \forall i=1,...,N.
    \label{eq. compression without privacy parallel}
  \end{align}
  
  Clearly, we have $R'_{p}(\bar{\gamma}, L^*(\bar \gamma)) \geq R'(\bar{\gamma}, L^*(\bar \gamma))$.
  To show that $R'(\bar{\gamma}, L^*(\bar \gamma)) \geq R'_{p}(\bar{\gamma}, H(S))$, we need to prove that parallel privatization is optimal for the following problem
  \begin{align}
    \notag R'(\bar{\gamma}, H(S)) = & \min_{p_{Y|X}} I(X;Y)\\
    \notag& \mbox{s.t.}~~ I(C_k;Y) \geq \gamma(C_k), \forall k=1,...,K, \\
      & ~~~~~~  I(X_i;Y_i) \geq \tau_i, \forall i=1,...,N. 
    \label{eq. compression without privacy}
  \end{align}
  
  \begin{Lem}\label{Lem.compression independently}
    Problem \eqref{eq. compression without privacy} and Problem \eqref{eq. compression without privacy parallel} have the same minimum value, i.e., $R'_{p}(\bar{\gamma}, H(S)) = R'(\bar{\gamma}, H(S))$.
    \end{Lem}
    \begin{IEEEproof}
This can be proved by showing that for any feasible privatization $p_{Y|X}(y|x)$ in \eqref{eq. compression without privacy}, there exists parallel privatization $p_{Y'|X}(y'|x)= \prod_i^N p_{Y'_i|X_i}(y'_i|x_i)$ which achieves the same released rate and satisfies all the constraints. Detail can be found in Appendix \ref{sec. proof of parallelized compression}.
    \end{IEEEproof}
    
  By Lemma~\ref{Lem.compression independently}, we now have 
\[
R'(\bar{\gamma}, L^*(\bar \gamma)) \geq R'(\bar{\gamma}, H(S)) = R_{p}'(\bar{\gamma}, H(S)).
\] 
Due to the restriction of using parallel privatization, both \eqref{eq. compression with parallelized} and \eqref{eq. compression without privacy parallel} can be decomposed and turned into an allocation problem as follows (by the same argument in Section \ref{subsec. optimal weighting}):
  \begin{align}
    \notag  R'_{p}(\bar{\gamma}, L^*(\bar \gamma)) =&\min_{\{\alpha_i\}_{i=1}^{N}}~ \sum_{i=1}^{N} \alpha_i\\
      \notag &\mbox{~~s.t.~}~ \sum_{i: X_i \in C_k}\alpha_i \geq \gamma(C_k), ~~\forall k=1,...,K,\\
     \notag & ~~~~~~~~ \alpha_i \geq \tau_i, ~~\forall i =1,...,N,\\
\notag      & ~~~~~~~~ \sum_{i=1}^N\alpha_i \leq L^*(\bar \gamma),
   \end{align}
  and 
  \begin{align}
    \notag  R'_{p}(\bar{\gamma}, H(S)) =&\min_{\{\alpha_i\}_{i=1}^{N}}~ \sum_{i=1}^{N} \alpha_i\\
      \notag &\mbox{~~s.t.~}~ \sum_{i: X_i \in C_k}\alpha_i \geq \gamma(C_k), ~~\forall k=1,...,K,\\
\notag      & ~~~~~~~~ \sum_{i=1}^N\alpha_i \leq L^*(\bar \gamma).
   \end{align}
Observing that the optimal allocation $\{\alpha_i\}_{i=1}^N$ in \eqref{eq. allocation problem 2} is actually the solution to both the above problems, we have 
  \begin{align*}
    L^*(\bar \gamma) + H(X)-H(S) &= R_{p}'(\bar{\gamma}, L^*(\bar \gamma)) \\
    &\geq R'(\bar{\gamma}, L^*(\bar \gamma))\\
     &\geq R'_{p}(\bar{\gamma}, H(S))\\
     &= L^*(\bar \gamma) + H(X)-H(S). 
  \end{align*}
  Since the minimum released rate is achieved by parallel privatization, the proof is complete.
\end{IEEEproof}

As a closing remark, since $R'(\bar{\gamma}, L^*(\bar \gamma))$ is an upper bound of $R(\bar{\gamma}, L^*(\bar \gamma))$, the rate given in Theorem~\ref{Lem. min privatization rate} serves as a sufficient condition of released rate to achieve the optimal privacy-utility tradeoff in \eqref{eq.mutual information opt problem}.

\section{Conclusions}\label{sec. conclusions}
In this paper, we examine the privacy-utility tradeoff with multiple possible tasks. 
Due to the lack of information of which task to be carried out, a robust privatization based on a given set of possible tasks is considered. We first derive the single-letter characterization of the optimal privacy-utility tradeoff. By applying log-loss distortion as the utility metric, the minimum privacy leakage problem is formulated as a compound version of the privacy funnel problem. Under the assumption that the raw data comprises multiple independent components and the private feature is a component-wise deterministic function of the raw data, we show that the minimum privacy leakage problem can be decomposed into multiple parallel privacy funnel problems with corresponding weightings, each of  which represents the amount of released information of each component of the raw data. We further solve the privacy funnel problem and show that the minimum leakage problem is equivalent to a linear program. The sufficient released rate to achieve the minimum privacy leakage is also studied. Numerical results are provided to illustrate the robustness of our approach as well as the impact of the selection of the set of possible tasks on the privacy-utility tradeoff. 

As for future work, we conjecture that parallel privatization remains optimal for the minimum released rate to attain the minimum privacy leakage subject to utility constraints. Extensions to non-deterministic private features and more general tasks are also promising directions to pursue. The multiple-tasks compound formulation may be extended from a finite set of tasks to a more general one. For example, the possible set of tasks may be described by all the tasks within a fixed distance of a given certain task instead of given specific multiple tasks, where the distance can be measured by KL divergence of other reasonable metrics. We believe that these information theoretical results will shed light on the practical implementation of robust privatized data release systems. In particular, the mutual information terms may be replaced by the variational lower and upper bounds approximated by empirical estimates, leading to a machine learning framework that trains a data-driven privatization mechanism.

\appendix
\subsection{Proof of Lemma \ref{Lem. RLD region}} \label{sec. proof of RLD region}
This lemma can be proved by the standard method used in rate-distortion theory \cite{Book_CoverElements} for the extension with multiple decoders. The additional leakage constraint can be dealt with by a method similar to that in \cite{Kittichokechai2016Privacy}. 

We first prove the achievability part. Assume 
\[
p_{Y,\hat{C}_1,...\hat{C}_K|X}(y,\hat{c}_1,...,\hat{c}_k|x)
\]
is the conditional probability mass function that satisfies \eqref{eq. leakage condition for region} and \eqref{eq. distortion condition for region}. We can construct the random codebook consisting of $2^{n(I(X;Y)+\delta)}$ sequences $y^n(v)$, $v \in [1:2^{n(I(X;Y)+\delta)}]$. Each sequence $y^n(v) = [y(1,v),...,y(n,v)]$ is generated independently according to $\Pi_{i=1}^n p_{Y}(y(i,v))$. We encode the raw data $x^n$ by $\phi^{(n)}(x^n) = v$ where $v$ is the index such that $(x^n, y^n(v))$ is jointly typical. The decoder of each task is given by $\theta^{(n)}(v) = \hat{C}^{(n)}_k(y^n(v)) \triangleq [\hat{C}_k(y(i, v)),...,\hat{C}_k(y(n, v)))]$, $\forall k=1,...,K$, where each element $\hat{C}_k(y(i,v))$ is a random mapping based on the conditional probability $p_{\hat{C}_k|Y}$. Clearly, we have $\frac{1}{n}\log|\mathcal{V}|\leq R + \delta$. Then, we show that such random coding achieves constraint \eqref{eq. leakage constraint} and \eqref{eq. distortion constraint}. 

For each $d_k(\cdot,\cdot)$, we can introduce the corresponding distortion function $d'_k(x, \hat{c}_k)\triangleq \E_{C_k}[d_k(C_k,\hat{c}_k)|X=x]$ such that $\E_{X,\hat{C}}[d'_k(X,\hat{C}_k)] = \E_{C_k,\hat{C}_k}[d_i(C_k,\hat{C}_k)]$. The expected distortion over all random codebooks can be rewritten as
\begin{align}
\notag \E[d_k(C_k^n,\theta^{(n)}(V))] &=\E[d_k(C_k^n,\hat{C}_k^{(n)}(Y^n(V)))]\\
&=\E[d_k'(X^n,\hat{C}^{(n)}_k(Y^n(V)))]. \label{eq. distortion d'}
\end{align}
Let $\mathcal{E}_k$ be the event where $(X^n, \hat{C}_k^{(n)}(Y^n(V)))$ are not jointly typical. Following the similar argument in the standard rate-distortion theory as treated in \cite{Book_ElGamalNetwork}, Equation \eqref{eq. distortion d'} can be bounded as follows:
\begin{align}
  \notag &\Pr(\mathcal{E}^c) \E[d_k'(X^n,\hat{C}^{(n)}_k(Y^n(V)))| \mathcal{E}^c] + \Pr(\mathcal{E})d'_{k,max} \\
\notag  &=  \Pr(\mathcal{E}^c)\E[d_k'(X^n,\hat{C}^{(n)}_k(Y^n(V)))| \mathcal{E}^c] + \delta_n \\
  &\leq  \Pr(\mathcal{E}^c)(1+\epsilon) \E[d_k'(X,\hat{C}_k)] + \delta_n \label{eq. typical set property}\\
\notag  &=  \Pr(\mathcal{E}^c)(1+\epsilon) \E[d_i(C,\hat{C}_k)] + \delta_n \\
\notag  &\leq  D_k + \delta_n,
\end{align}
where $d'_{k,max}$ is the maximum value of $d'_k(\cdot, \cdot)$ and $\delta_n \rightarrow 0$ as $n \rightarrow \infty$. By the law of large number and the property of jointly typical set, we have $\Pr(\mathcal{E}) \rightarrow 0$ as $n\rightarrow \infty$. The inequality in \eqref{eq. typical set property} comes from  the typical average lemma \cite{Book_ElGamalNetwork}. 

For the privacy leakage, we have
\begin{align}
\notag &I(S^n; V) \\
\notag  &= H(S^n) - H(S^n|V)\\
\notag  & =H(S^n) - H(S^n, X^n|V) + H(X^n|V,S^n)\\
\notag  &\leq H(S^n) - H(S^n, X^n) + H(V) + H(X^n|V,S^n) \\
  & = - nH(X| S) +n(I(X; Y) + \delta_n) + H(X^n|V,S^n) \label{eq. iid and data size}\\ 
  & \leq n( I(X; Y) - H(X|S) + \delta_n + H(X| Y,S) +\delta_n') \label{eq. apply Cairs lemma} \\
\notag  &  = n\left(-H(X|S) + H(Y) - H(Y|X,S) + H(X| Y,S) +\delta''_n \right)  \\
\notag  & = n(I(S;Y) +\delta''_n)
\leq n(L + \delta''_n).
  \end{align}
The equality in \eqref{eq. iid and data size} comes from the fact that $(S^n, X^n)$ are i.i.d. and $|V| \leq 2^{ n(I(X;Y)+\delta)}$. The inequality in \eqref{eq. apply Cairs lemma} follows the result in \cite[Lemma 2]{Kittichokechai2016Privacy} which states that if $\Pr((X^n,Y^n(V),S^n) \in \mathcal{A}^{(n)}) \rightarrow 1$, where $\mathcal{A}^{(n)}$ is the typical set, then  $H(X^n|V,S^n) \leq n(H(X| Y,S) + \delta_n')$. Thus, we complete the proof of achievability part.
  
Next, we prove the converse part. Let us consider an achievable tuple $(R,L,D_1,...,D_K)$. There exists a sequence of $(|\mathcal{V}|,n)$-scheme with encoder $\phi^{(n)}(X^n) = V$ and decoders $\theta_1^{(n)},...\theta^{(n)}$ that satisfy \eqref{eq. rate constraint}, \eqref{eq. leakage constraint}, and \eqref{eq. distortion constraint}. Hence, for the rate constraint, 
\begin{align}
\notag    n(R+\delta) & \geq H(V)\\
\notag    &\geq I(V;X^n)\\
\notag    &=\sum_{i=1}^n I(X(i);V|X(1),...,X(i-1))\\
    &=\sum_{i=1}^n I(X(i);V,X(1),...,X(i-1)) \label{eq. converse pf rate indep property}\\
    &\geq \sum_{i=1}^n I(X(i);V) \label{eq. converse proof compression rate}.
  \end{align}
  The equality in \eqref{eq. converse pf rate indep property} follows the independence of $X(i)$. Similarly, for the leakage constraint, we have
\begin{align}
\notag    n(L+\delta) &\geq I(S^n;V)\\
\notag    &=\sum_{i=1}^n I(S(i);V|S(1),...,S(i-1))\\
    &\geq \sum_{i=1}^n I(S(i);V). \label{eq. converse proof leakage}
  \end{align}
For the distortion constraints, we have
\begin{align}
\notag    n(D_k+\delta) &\geq \E[d^{(n)}(C_k^n,\theta_k^{(n)}(V))]\\
  & = \sum_{i=1}^n \E[d_k(C_{k}(i), \theta_{k}(i, V))], \label{eq. converse proof distortion}
\end{align}
where $\theta_{k}(i, V)$ denotes the $i$-th element of $\theta^{(n)}_{k}(V)$.
Let $Y(i) = V$ and $\hat{C}_{k}(i) =\theta_{k}(i,V)$, $\forall i=1,...,n$ and $\forall k=1,...,K$. By introducing the time-sharing random variable $Q = i \in \{1,...,n\}$ with equal probability,  we can rewrite the inequalities in \eqref{eq. converse proof compression rate}, \eqref{eq. converse proof leakage} and \eqref{eq. converse proof distortion}  as
\begin{align*}
    R +\delta  &\geq \frac{1}{n}\sum_{i=1}^n I(X(i);Y(i))\\
      &= I(X_Q;Y_Q|Q) = I(X_Q;Y_Q, Q),\\
    L +\delta  &\geq \frac{1}{n}\sum_{i=1}^n I(S(i);Y(i))\\
      &= I(S_Q;Y_Q|Q) = I(S_Q;Y_Q, Q),\\
    D_k +\delta &\geq \frac{1}{n} \sum_{i=1}^n \E[d_k(C_{k}(i), \hat{C}_{k}(i))]\\
      &= \E[d_k(C_{k,Q}, \hat{C}_{k,Q})].
\end{align*}
Due to the i.i.d. property, $X_Q$, $S_Q$, $C_{k,Q}$ have the same distribution as $X$, $S$, $C_k$, respectively. Now, select $Y=(Y_Q, Q)$ and  $\hat{C}_{k}=\hat{C}_{k,Q}$, $\forall k=1,...,K$ as the desired random variables to achieve \eqref{eq. rate condition} -- \eqref{eq. distortion condition for region}. Observing that $(S(i), C_{1}(i),...,C_{K}(i))-  X(i) -  Y(i) -  (\hat{C}_{1}(i),...,\hat{C}_{K}(i))$ forms a Markov chain, we have  $(S, C_{1},...,C_{K})-  X -  Y -  (\hat{C}_{1},...,\hat{C}_{K})$, which completes the proof of the converse part. Finally, the cardinality bound can be proved by using the supporting lemma \cite{Book_ElGamalNetwork}.

\begin{figure*}[!t]
\normalsize
\begin{align*}
&\E_{Y, \hat{C}_K} \E[d_k(C_k, \hat{C}_k)|Y, \hat{C}_k]\\
&= \sum_{y, \hat{c}_k} p_{Y, \hat{C}_k}(y, \hat{c}_k) \sum_{c_k} p_{C_k|Y,\hat{C}_k}(c_k|y, \hat{c}_k) \left(\log\frac{p_{C_k|Y,\hat{C}_k}(c_k|y, \hat{c}_k)}{\hat{c}_k(c_k)}+ \log \frac{1}{p_{C_k|Y,\hat{C}_k}(c_k|y, \hat{c}_k )}\right)\\
&= \E_{Y,\hat{C}_k} \left[ D_{KL}(p_{C_k|Y,\hat{C}_k}\Vert\hat{C}_k)\right] + H(C_k| Y, \hat{C}_k), 
\end{align*}
\hrulefill
\vspace*{4pt}
\end{figure*}

\subsection{Proof of the equivalence between \eqref{eq. distortion condition for region} and \eqref{eq. condition entropy constraint} under the log-loss distortion} \label{sec. proof of equivalent of distortion constraint}
The distortion constraints \eqref{eq. distortion condition for region} can be written as shown on the top of the next page, where $D_{KL}(\cdot\Vert\cdot)$ is the Kullback-Leibler (KL) divergence. Since KL divergence is non-negative and due to the Markov chain property, we immediately have the converse part, that is, $D_k \geq \E[d_k(C_k, \hat{C}_k)]$ implies $D_k \geq  H(C_k| Y)$. For the achievability part, we can always select $\hat{C}_k(c_k) = h_k(Y, c_k)$ where $h_k(y, c_k) = p_{C_k|Y}(c_k|y)$ such that $\E[d_k(C_k, \hat{C}_k)] =H(C_k| Y)$. Since $\hat{C}_k$ is a deterministic function of $Y$, the Markov chain still holds. Hence, we have  $D_k \geq H(C_k| Y)$ implies $D_k \geq \E[d_k(C_k, \hat{C}_k)]$. Now, let $\gamma(C_k) = H(C_k) - D_k$, we can replace the distortion constraints \eqref{eq. distortion condition for region} by the mutual information based constraint \eqref{eq. condition entropy constraint}. 

\subsection{Proof of Lemma \ref{Lem.compression independently}} \label{sec. proof of parallelized compression}
We prove this lemma by showing that for any feasible privatization $Y$ in \eqref{eq. compression without privacy}, there exists a parallel privatization $Y' = [Y'_1, Y'_2, ..., Y'_N]$ which satisfies 
\begin{align*}
&p_{Y'|X}(y'|x) = \prod_{i}^{N} p_{Y'_i|X_i}(y'_i|x_i),\\
&I(X;Y') = I(X;Y),\\
&I(C_k;Y') \geq I(C_k;Y) , ~~~\forall k=1,...,K,\\
&I(X_i;Y') \geq I(X_i;Y) , ~~~\forall i=1,...,N.
\end{align*}

Let the alphabet $\mathcal{Y}'_i= \mathcal{X}_1\times...\times \mathcal{X}_{i-1}\times \mathcal{Y}$ and denote the realization by $y'_i=(x_1^{(i)},...,x_{i-1}^{(i)},y^{(i)})$, $\forall i$. We construct a parallel $Y'$ by $p_{Y'|X}(y'|x) = \prod_{i}^{N} p_{Y'_i|X_i}(y'_i|x_i)$ with $p_{Y'_i|X_i}(y'_i|x_i) = p_{X_1,...,X_{i-1},Y|X_i}(x_1^{(i)},...,x_{i-1}^{(i)},y^{(i)}|x_i)$. 
Since $\{X_i,Y_i'\}_{i=1}^{N}$ are independent, we have 
\begin{align*}
    H(X|Y') &= \sum_{i=}^N H(X_i|Y_i')\\
    & =\sum_{i=1}^N H(X_i|X_1,...,X_{i-1},Y)\\
    & = H(X|Y).
\end{align*}
For the utility constraints, we have
\begin{align*}
    I(C_k;Y') &= H(C_k) - H(C_k|Y')\\
    & = H(C_k) - \sum \limits_{i: X_i \in C_k} H(X_i|Y_i')\\
    & = H(C_k) - \sum \limits_{i: X_i \in C_k} H(X_i|X_1,...X_{i-1},Y)\\
    & \geq H(C_k) - \sum \limits_{i: X_i \in C_k} H(X_i|\{X_1,...X_{i-1}\}\cap C_k,Y)\\
    & = H(C_k) - H(C_k|Y) = I(C_k;Y).
\end{align*}
Finally, for the individual constraint, we have
\begin{align*}
  H(X_i|Y') &= H(X_i|Y_i')\\
  & = H(X_i|X_1,...,X_{i-1},Y)\\
  & \leq  H(X_i|Y),
\end{align*}
which completes the proof.

\bibliographystyle{IEEEtran}

\begin{IEEEbiographynophoto}{Ta-Yuan Liu}
	received his B.S. degree in Electrical Engineering and his Ph.D. degree in Communications Engineering from National Tsing Hua University, Hsinchu, Taiwan, in 2009 and 2016 respectively. In 2013, he served as a Visiting Student at the University of Maryland, USA. From 2016 to 2019, he worked as a wireless system engineer at Realtek Semiconductor Corp., Hsinchu, Taiwan. Following that,  he held a Post-Doctoral Research position at National Taiwan University, Taipei, Taiwan from 2019 to 2021. Currently he is employed as a Senior Engineer at Mediatek Inc., Hsinchu, Taiwan. 
He received the best student paper award from National Symposium on Telecommunication (NST) in 2011 and the Poster Runner-up from Croucher Summer Course in Information Theory (CSCIT) 2021. His research interests include physical layer security, data privacy preserving, AI-native wireless systems, and information theory.
\end{IEEEbiographynophoto}

\begin{IEEEbiographynophoto}{I-Hsiang Wang}
	received the B.Sc. degree in Electrical Engineering from National Taiwan University, Taiwan, in 2006. He received a Ph.D. degree in Electrical Engineering and Computer Sciences from the University of California at Berkeley, USA, in 2011. From 2011 to 2013, he was a postdoctoral researcher at \`{E}cole Polytechnique F\`{e}d\`{e}rale de Lausanne, Switzerland. Since 2013, he has been at the Department of Electrical Engineering in National Taiwan University, where he is now a professor. His research interests include network information theory, networked data analysis, and statistical learning. He was a finalist of the Best Student Paper Award of IEEE International Symposium on Information Theory, 2011. He received the 2017 IEEE Information Theory Society Taipei Chapter and IEEE Communications Society Taipei/Tainan Chapters Best Paper Award for Young Scholars.
\end{IEEEbiographynophoto}

\end{document}